\documentclass[a4paper,12pt]{article}

\usepackage[margin=24mm]{geometry}
\usepackage[T1]{fontenc}
\usepackage{lmodern,microtype}
\usepackage{amsmath,amssymb,mathtools,bm}
\usepackage{booktabs,tabularx,array}
\usepackage{graphicx,float}
\usepackage[dvipsnames]{xcolor}
\usepackage[colorlinks=true,linkcolor=MidnightBlue,citecolor=MidnightBlue,urlcolor=BrickRed]{hyperref}
\usepackage[nameinlink,noabbrev]{cleveref}
\usepackage{authblk}
\usepackage{enumitem}
\usepackage{todonotes}
\allowdisplaybreaks
\numberwithin{equation}{section}

\newcommand{\gev}{\text{GeV}\,}

\title{Probing Compressed Dark Sectors at a Muon Collider through Dilepton Signatures and Machine Learning}

\author[1]{Mohamed Belfkir\thanks{
\href{mailto:mohamed.belfkir@udst.edu.qa}
{\textcolor{MidnightBlue}{\nolinkurl{mohamed.belfkir@udst.edu.qa}}}
}}

\author[2]{Ayoub Hmissou\thanks{
\href{mailto:ayoub1hmissou@gmail.com}
{\textcolor{MidnightBlue}{\nolinkurl{ayoub1hmissou@gmail.com}}}
}}

\author[2]{Larbi Rahili\thanks{
\href{mailto:rahililarbi@gmail.com}
{\textcolor{MidnightBlue}{\nolinkurl{rahililarbi@gmail.com}}}
}}

\affil[1]{University of Doha for Science and Technology (UDST), Doha, Qatar}

\affil[2]{Laboratory of Theoretical and High Energy Physics,
Faculty of Science, Ibnou Zohr University,
B.P. 8106, Agadir, Morocco}

\date{}

\begin{document}
\maketitle

\begin{abstract}
The inert doublet model (IDM), a minimal extension of the Standard Model (SM), provides a scalar dark matter (DM) candidate that belongs to the additional Higgs doublet. The model faces challenges in achieving the correct relic abundance for compressed spectra and DM masses in the high-mass range. In this work we investigate the prospects of detecting dark matter candidates at future high-energy muon colliders. Focusing on the process $\mu^+ \mu^- \to AH$ with subsequent decay $A \to H Z^*$ and $Z^* \to \ell^+\ell^-$, we explore the sensitivity of machine learning techniques, specifically Boosted Decision Trees (BDT) and XGBoost, to distinguish the signal from Standard Model backgrounds. We demonstrate that the combination of advanced multivariate analysis with high-luminosity muon collisions enhances the discovery potential for inert scalars in compressed mass spectra scenarios.
\end{abstract}

\clearpage
\tableofcontents

\section{Introduction}\label{sec:intro}

A wide range of astrophysical and cosmological observations, from galactic rotation curves and gravitational lensing to the anisotropies of the cosmic microwave background, establishes that about $85\%$ of the matter content of the Univers is non-baryonic dark matter (DM), with a present-day abundance $\Omega_{\rm DM}h^2 \simeq 0.12$~\cite{Bertone:2004pz,Planck:2018vyg}. The Standard Model (SM) of particle physics provides no particle able to play this role, and the identification of the DM constituent is therefore one of the most compelling indications of physics beyond the SM (BSM). Among the manu proposed candidates, weakly interacting massive particles (WIMPs) remain particularly attractive, since thermal freeze-out with electroweak-strength interactions naturally produces a relic density of the observed order~\cite{Arcadi:2017kky}. Their simplest realisations, however, are increasingly constrained by direct-detection (DD) experiments, most recently by LUX-ZEPLIN (LZ)~\cite{LZ:2024zvo}. This situation has shifted attetion towards scenarios i which the DM couples feebly to nucleons but retains sizeable electroweak gauge interactions, for which high-energy colliders provide a complementary and often decisive probe.

The inert doublet model (IDM), firstly suggested in~\cite{Deshpande:1977rw} and widely investigated in the literature~\cite{LopezHonorez:2006gr,Arhrib:2015hoa,Belyaev:2016lok,Abouabid:2023cdz,Arhrib:2026xhv}, is one of the most economical and predictive frameworks of this type. It extends the SM scalar sector by a second $SU(2)_{L}$ doublet that is odd under un exact $\mathbb{Z}_2$ symmetry and acquires no vacuum expectation value. As a result, the inert doublet has no Yukawa couplings to SM fermions, and its interactions with the SM proceed only through the electroweak gauge bosons and the scalar potential. The physical spectrum contains two neutral states, the CP-even $H$ and the CP-odd $A$, in addition to a charged pair $H^\pm$; the lightest neutral state, taken here to be $H$, is stable and constitutes a DM candidate. Phenomenologically, the DM within the model has been studied extensively~\cite{LopezHonorez:2010tb,Goudelis:2013uca,Arhrib:2013ela,Hambye:2009pw}, and three regimes can be distinguished depending on the $H$ mass. However, such a compressed spectra is notoriously difficult to probe at hadron colliders. Indeed, when the mass splitting $\Delta m_{AH} \equiv m_A-m_H$ is smaller than $m_Z$, the decay $A \to H Z^{*} \to H \ell \bar{\ell}$ produces soft visible particles accompanied by missing momentum, which are hard to trigger on and are overwhelmed by SM backgrounds. 

LEP excluded a part of the light-mass region~\cite{Lundstrom:2008ai}, and LHC searches in dilepton, mulilepton and vector boson fusion (VBF) final states~\cite{Belanger:2015kga,Dercks:2018wch}, as well as dedicated soft-lepton analyses designed for compressed electroweak spectra~\cite{ATLAS:2019lng}, have reduced sensitivity to inert scalars with masses above the electroweak scale. Lepton colliders, with a well-defined initial state, low QCD activity and the possibility of reconstructing the event kinematics from the known collision energy, are much better suited to this task, as demonstrated by previous studies of the IDM at future $e^+e^-$ facilities~\cite{Kalinowski:2018ylg,Kalinowski:2018kdn}.

Also in this context, a multi-TeV muon collider (MuC) combines the clean environment of a lepton collider with a center-of-mass energy that is otherwise accessible only at hadron machines~\cite{Delahaye:2019omf,AlAli:2021let,Accettura:2023ked}. Additionally, two features make it particularly suited to the inert sector. First, the electroweak pair production $\mu^+\mu^- \to H A$ proceeds through $s$-channel $Z$ exchange with a coupling fixed entirely by gauge invariance, so that its rate depends only on the masses of the produced states. Second, at multi-TeV energies the incoming muons act as an intense source of electroweak gauge bosons, and VBF processes such as $W^+W^- \to H A$ become important and can even dominate over $s$-channel annihilation~\cite{Costantini:2020stv}. In this study, we focus on the compressed regime and on the soft-dilepton final state $\mu^+\mu^- \to H A$, followed by the $A \to H Z^{*} \to H \ell^+ \ell^-$ ($\ell=e,\mu$), together with the complementary channel $\mu^+\mu^- \to H A\, \nu_\mu \bar\nu_\mu$, which contains the $W^+W^-$-fusion topologies and is computed from the complete gauge-invariant matrix element. Both channels yield an opposite-sign, same-flavour lepton pair with missing momentum, whose invariant mass is bounded from above by $m_{\ell\ell} \le \Delta m_{AH}$. The kinematic endpoint allows the mass splitting in the inert spectrum to be determined.

Because the signal leptons are soft and the expected rates are at the level of a fraction of a femtobran once the leptonic branching fraction and the fiducial selection requirements are taken into account, the sensitivity depends critically on how efficiently the signal can be separated from the SM backgrounds. Machine-learning methods, and boosted decision trees (BDTs) in particular, are now standard tools for this purpose in collider physics~\cite{Guest:2018yhq,Feickert:2021ajf}. In this work, we first construct a transparent cut-based selection, which serves as a physical baseline, and then develop a multivariate strategy build on gradient-boosted decision trees implemented with XGBoost~\cite{Chen:2016btl}. A first classifier, parametrised in terms of $m_H$ and $\Delta m_{AH}$~\cite{Baldi:2016fzo}, discriminates the signal from the backgrounds over the whole set of benchmark hypotheses and is shown to interpolate reliably to hypotheses excluded from its training. A second classifier separates VBF-like from $s$-channel-like signal events and defines two orthogonal analysis categories, so that the different signal-to-background compositions of the two production topologies are preserved. The parameterized classifier output is then used in a binned profile-likelihood analysis~\cite{Read:2002hq} implemented with \texttt{pyhf}~\cite{Heinrich:2021gyp}, from which we derive expected $95\%$ confidence level (C.L.) upper limits on the signal cross section.

The paper is organised as follows. In section~\ref{sec:model} we review the IDM and the theoretical and experimental constraints imposed on its parameter space. Section~\ref{sec:numerical} presents the parameter space, the production cross sections in the direct and VBF-enhanced channels, and the benchmark points. The cut-based collider analysis is described in section~\ref{sec:cutbased}, and the multivariate strategy in section~\ref{sec:mva}. The statistical analysis and our results are presented in section~\ref{sec:stat}, and we conclude in section~\ref{sec:conclusion}.


\section{Model}\label{sec:model}
The IDM extends the SM scalar sector through the introduction of a second $SU(2)_L$ doublet, $\Phi_2$, carrying hypercharge $Y=1/2$. The model is endowed with an exact discrete $\mathbb{Z}_2$ symmetry, under which the scalar doublets transform as
\begin{equation}
\mathbb{Z}_2:\qquad \Phi_1\longrightarrow\Phi_1,
\qquad \Phi_2\longrightarrow-\Phi_2,
\label{eq:z2-symmetry}
\end{equation}
while all SM fields are $\mathbb{Z}_2$-even. This symmetry is preserved after electroweak symmetry breaking, with the inert doublet acquiring no vacuum expectation value, $\langle\Phi_2\rangle=0$. Accordingly, $\Phi_2$ does not mix with the SM Higgs doublet, and renormalisable Yukawa interactions between $\Phi_2$ and SM fermions are forbidden. In the electroweak broken phase, the two doublets can be written as
\begin{equation}
\Phi_1 =
\begin{pmatrix}
G^+ \\[1mm]
(v+h+iG^0)/\sqrt{2}
\end{pmatrix},
\qquad
\Phi_2 =
\begin{pmatrix}
H^+ \\[1mm]
(H+iA)/\sqrt{2}
\end{pmatrix}.
\label{eq:doublets}
\end{equation}
where $v\simeq246$ GeV denotes the electroweak vacuum expectation value, $G^0$ and $G^\pm$ are the would-be Goldstone bosons, and $h$ is identified with the observed SM-like Higgs boson. The physical inert sector therefore comprises four $\mathbb{Z}_2$-odd scalar states: the CP-even neutral scalar $H$, the CP-odd neutral scalar $A$, and the charged scalars $H^\pm$.

The most general renormalisable and CP-conserving scalar potential invariant under the gauge symmetry and the $\mathbb{Z}*2$ transformation defined in eq.~\eqref{eq:z2-symmetry} takes the form~\cite{Deshpande:1977rw,Barbieri:2006dq}

\begin{align}
V(\Phi_1, \Phi_2) &= \mu_1^2 |\Phi_1|^2 + \mu_2^2 |\Phi_2|^2 + \lambda_1 |\Phi_1|^4 + \lambda_2 |\Phi_2|^4 \nonumber \\
&\quad + \lambda_3 |\Phi_1|^2 |\Phi_2|^2 + \lambda_4 |\Phi_1^\dagger \Phi_2|^2 + \frac{\lambda_5}{2} \left[ (\Phi_1^\dagger \Phi_2)^2 + \text{h.c.} \right].
\label{eq:potentia}
\end{align}
where all parameters are taken to be real. At the inert vacuum, the minimisation condition yields $\mu_1^2=-\lambda_1v^2$, while the scalar masses at tree level are given by
\begin{eqnarray}
&&  m_h^2 = - 2 \mu_1^2 = 2 \lambda_1 v^2 \nonumber \\
&&  m_{H}^2 = \mu_2^2 + \frac{1}{2} (\lambda_3 + \lambda_4 + \lambda_5) v^2 = \mu_2^2 + \lambda_{L} v^2 \nonumber \\
&&  m_{A}^2 = \mu_2^2 +\frac{1}{2} (\lambda_3 + \lambda_4 - \lambda_5) v^2 = \mu_2^2 + \lambda_{S} v^2 \nonumber \\
&&  m_{H^{\pm}}^2 = \mu_2^2 + \frac{1}{2} \lambda_3 v^2.
\label{eq:masses}
\end{eqnarray}
with the coupling combinations
\begin{equation}
\lambda_L=\frac{\lambda_3+\lambda_4+\lambda_5}{2},
\qquad
\lambda_S=\frac{\lambda_3+\lambda_4-\lambda_5}{2}.
\label{eq:portal_couplings}
\end{equation}
The parameter $\lambda_L$ governs the tree-level $hHH$ interaction and is consequently of particular relevance to the relic abundance and direct-detection phenomenology of $H$. For the phenomenological study, we adopt an input parametrisation in terms of the electroweak vacuum expectation value, the physical scalar masses, and two independent quartic couplings,
\begin{equation}
\Omega=\{v,m_h,m_H,m_A,m_{H^\pm},\lambda_2,\lambda_L\}.
\label{eq:physical-basis}
\end{equation}
The corresponding expressions for the remaining inert-sector parameters are
\begin{equation}
\begin{aligned}
\mu_2^2
&= m_H^2-\lambda_L v^2,
\\[1mm]
\lambda_3
&= 2\lambda_L
+\frac{2\left(m_{H^\pm}^2-m_H^2\right)}{v^2},
\\[1mm]
\lambda_4
&= \frac{m_H^2+m_A^2-2m_{H^\pm}^2}{v^2},
\\[1mm]
\lambda_5
&= \frac{m_H^2-m_A^2}{v^2}.
\end{aligned}
\label{eq:quartics-physical}
\end{equation}
The lightest neutral scalar $H$ is stable due to the $\mathbb{Z}_2$ symmetry and constitutes a dark matter candidate. The phenomenology of the IDM depends on the mass splittings between $H$, $A$, and $H^\pm$, which control the kinematics of decays and the observability of the signal at colliders.

From a theoretical and experimental requirements point of view, the IDM as many extensions BSM, is subject to many constraints. We recall here:
\begin{enumerate}[label=(\alph*)]
 \item \textbf{Boundedness from below:} the quartic potential must remain positive in every large-field direction. At tree level this requires ~\cite{Deshpande:1977rw}
 \begin{equation}
  \lambda_1>0,\quad \lambda_2>0,\quad
  \lambda_3>-2\sqrt{\lambda_1\lambda_2},\quad
  \lambda_3+\lambda_4-|\lambda_5|>-2\sqrt{\lambda_1\lambda_2}.
  \label{eq:bfb}
 \end{equation}
 \item \textbf{Perturbative unitarity and the inert vacuum:} the eigenvalues of the tree-level scalar-scattering matrix are required to satisfy the standard perturbative-unitarity bounds, and the electroweak vacuum with $\langle\Phi_2\rangle=0$ must be the global minimum of the potential.~\cite{Espinosa:2011ax,Branchina:2018qlf}
 \item \textbf{Electroweak precision data:} the inert scalars contribute radiatively to the oblique parameters~\cite{Peskin:1990zt,Peskin:1991sw}, making this requirement particularly sensitive to splittings among $H$, $A$, and $H^\pm$. We set $U=0$ and use
 \begin{equation}
  S=-0.05\pm0.07,\qquad T=0.00\pm0.06,
  \qquad \rho_{ST}=0.93,
  \label{eq:st-inputs}
 \end{equation}
 retaining points with $\chi^2_{ST}<5.99$~\cite{ParticleDataGroup:2024cfk}.
 \item \textbf{Direct searches:} mass- and splitting-dependent exclusions from LEP searches for neutral and charged inert scalars are applied. In particular, we impose the approximate charged-scalar lower limit $m_{H^\pm}>70$ GeV rather than assigning a universal lower bound to the neutral states.
 \item \textbf{Higgs observables:} when kinematically accessible, the non-standard modes $h\to HH$, $AA$, and $H^+H^-$ are included in the Higgs width, and we require $\mathrm{BR}(h\to\mathrm{inv})<0.13$. The charged-scalar loop contribution to $h\to\gamma\gamma$ is also required to remain compatible with the measured inclusive Higgs rate~\cite{ParticleDataGroup:2024cfk}.
 \item \textbf{Dark-matter abundance:} the relic density is evaluated with \texttt{micrOMEGAs}~\cite{Alguero:2023zol} and imposed as an upper limit,
 \begin{equation}
  \Omega_Hh^2\leq \Omega_{\rm DM}^{\rm obs}h^2\simeq0.12,
  \label{eq:relic-upper}
 \end{equation}
 thereby allowing $H$ to account for only a fraction of the cosmological DM density.
 \item \textbf{Direct detection:} for an underabundant thermal component, the spin-independent nucleon-scattering cross section is rescaled by the abundance fraction,
 \begin{equation}
  \xi\,\sigma_{\rm SI}(m_H)\leq\sigma_{\rm SI}^{\rm LZ}(m_H),
  \qquad
  \xi=\frac{\Omega_Hh^2}{\Omega_{\rm DM}^{\rm obs}h^2},
  \label{eq:dd-rescaling}
 \end{equation}
 and compared with the LUX--ZEPLIN exclusion limit~\cite{LZ:2024zvo}.
\end{enumerate}

These requirements define the viable parameter space used in the production-level study of section~\ref{sec:numerical}. The additional compressed-spectrum condition relevant to the soft-dilepton signature is introduced separately in section~\ref{sec:numerical}.

\section{Numerical Results}
\subsection{Scan strategy}\label{sec:numerical}
We perform random scans over the inert doublet model (IDM) parameter space within the following ranges:
\begin{eqnarray}
	\label{Ei19}
	&& m_h = 125.09\,\gev, \hspace*{0.25cm} m_{H},\, m_{A} \in [65,\,10^3]\,\gev, \hspace*{0.25cm} m_{H^\pm}  \in [70,\,10^3]\,\gev, \nonumber\\
	&& \lambda_2 \in [0,\,6]\,, \hspace*{0.25cm} \lambda_{L} \in [-0.5,\,0.5],\,
\end{eqnarray}
All viable points are required to satisfy the theoretical and experimental constraints described in section ~\ref{sec:model}. We retain the hierarchy $m_H<m_A,m_{H^\pm}$, so that H is the stable Dark-Matter candidate, and focus on the compressed region:
\begin{equation}
 \Delta m_{AH}\equiv m_A-m_H<m_Z.
 \label{eq:compressed}
\end{equation}
The compressed spectrum is therefore a selection imposed on the viable parameter space, rather than a consequence of the constraints alone. In this section, we first examine the inclusive production rate before decays and event-selection cuts, and then identify the benchmark points used in the collider analysis.
\subsection{Direct associated production $\mu^+\mu^- \to H A $}\label{subsec:direct}
\begin{figure}[H]
	\centering\includegraphics[width=0.5\textwidth]{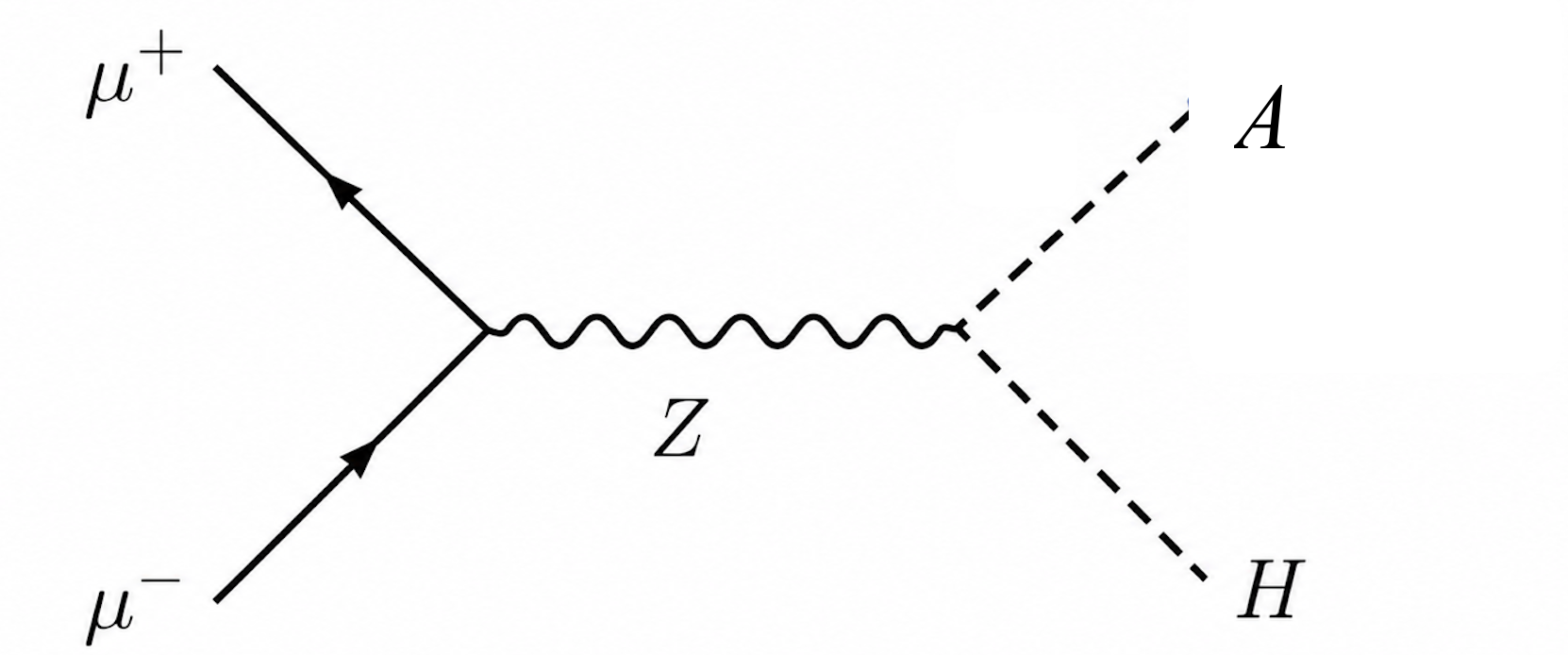}
    \caption{Tree level Feynman diagrams for $\mu^+ \mu^- \to H A $ } 
	\label{fig:fig1}
\end{figure}

\begin{figure}[H]
	\centering\includegraphics[width=0.5\textwidth]{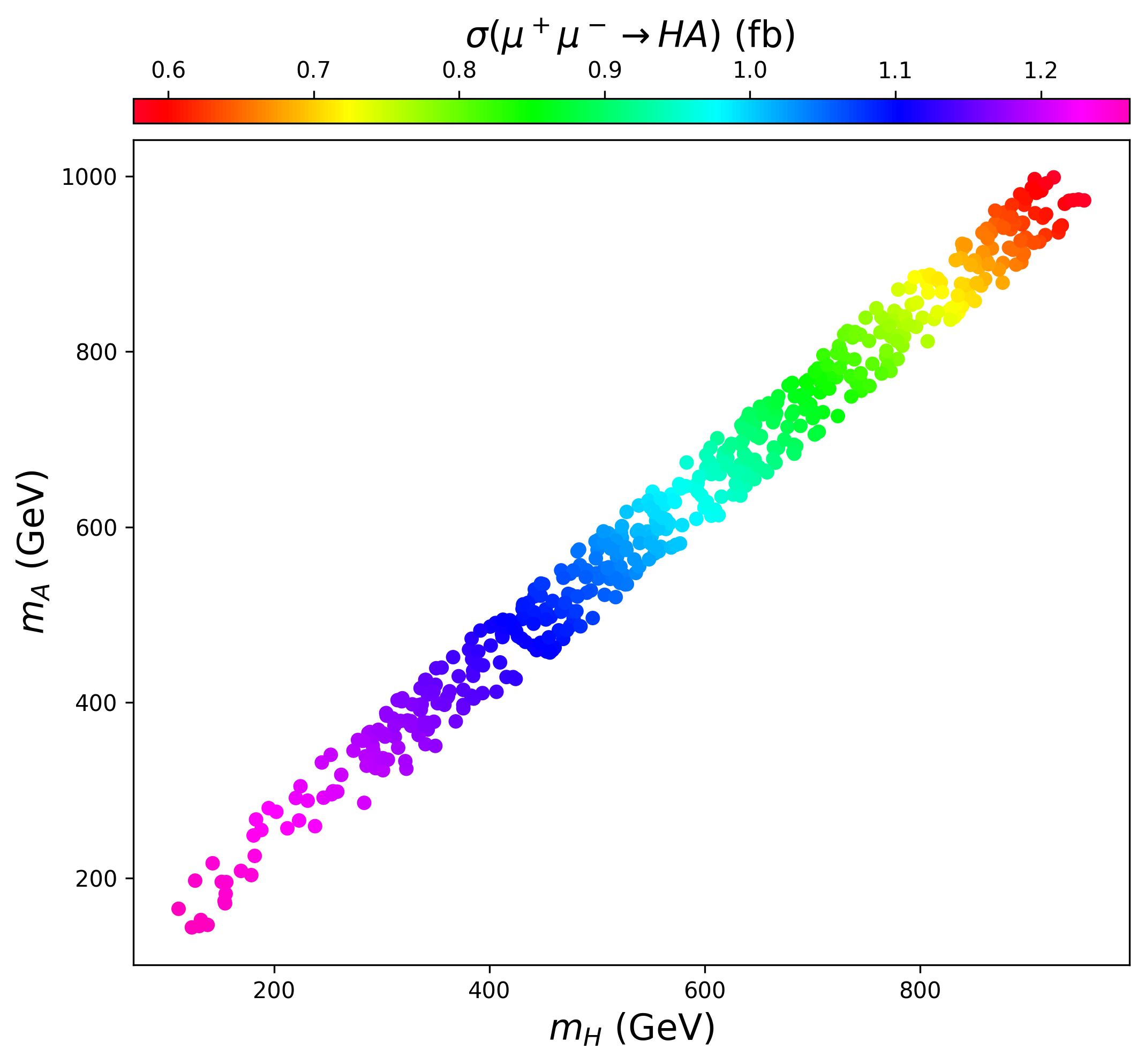}
	\caption{Correlation between the $m_H$ and $m_A$ with the cross section value represented by the color map} 
	\label{fig:fig2}
\end{figure}

\begin{figure}[H]
	\centering
	\includegraphics[width=0.4\textwidth]{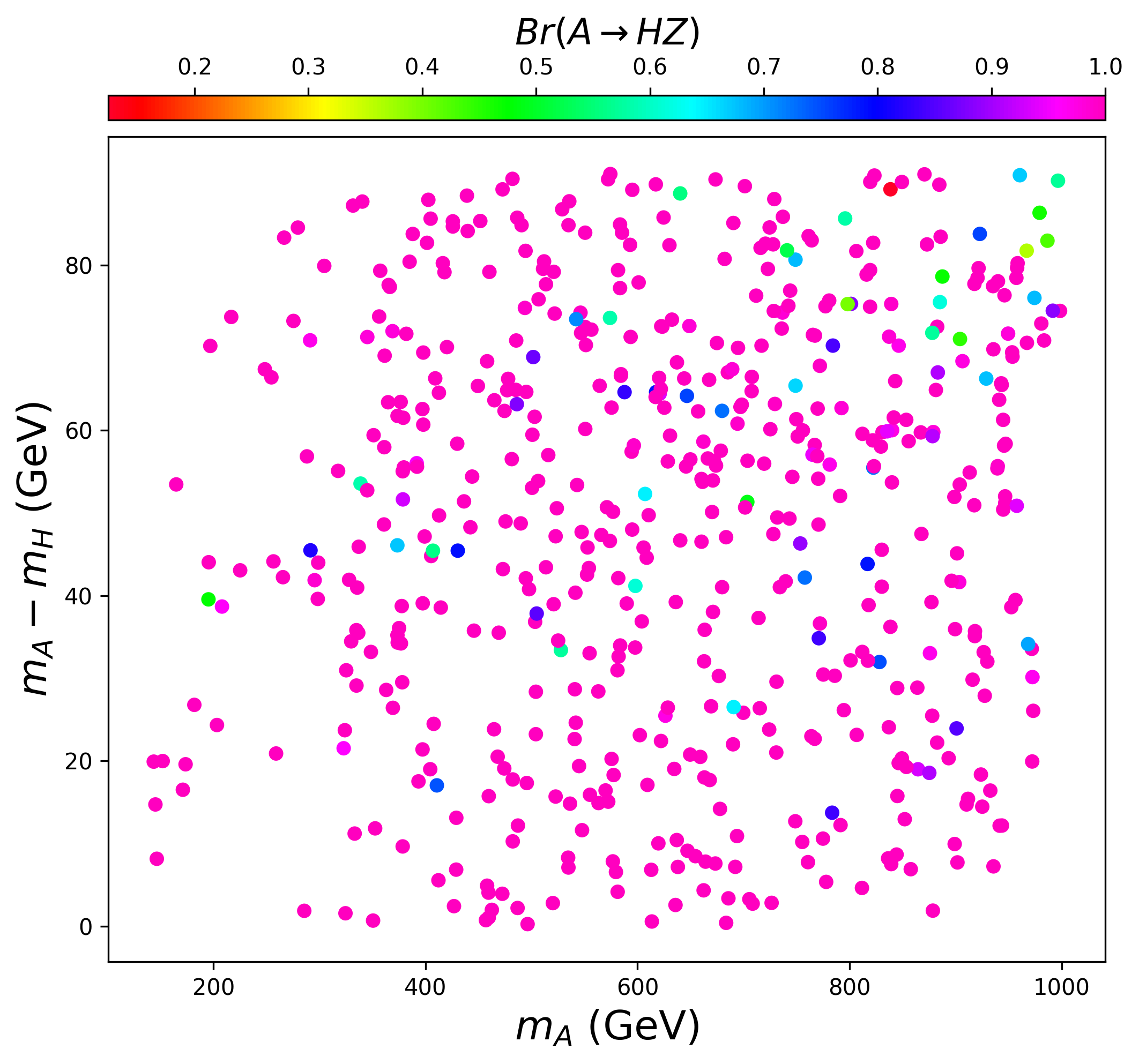}
	\includegraphics[width=0.4\textwidth]{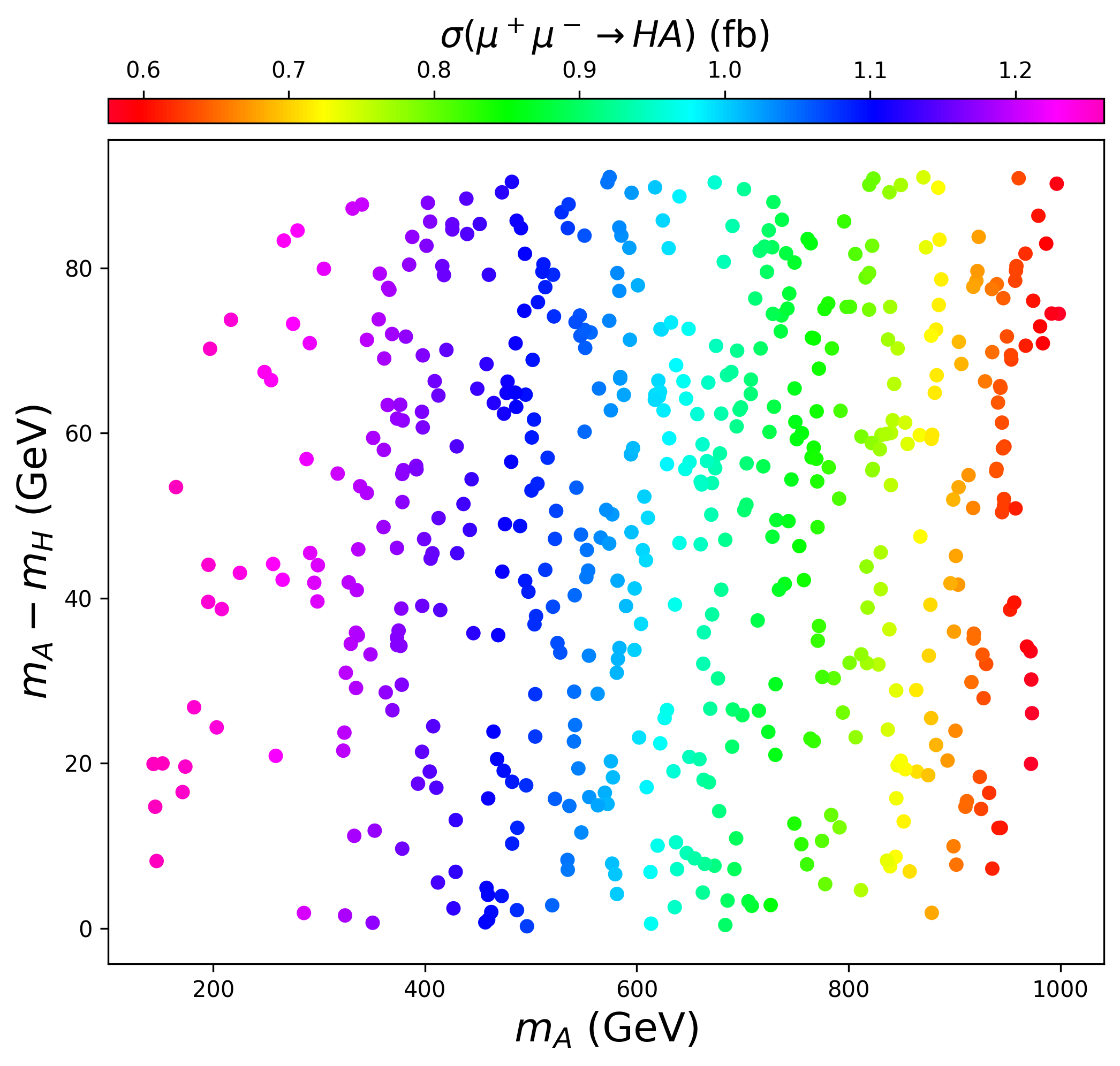}
	\includegraphics[width=0.4\textwidth]{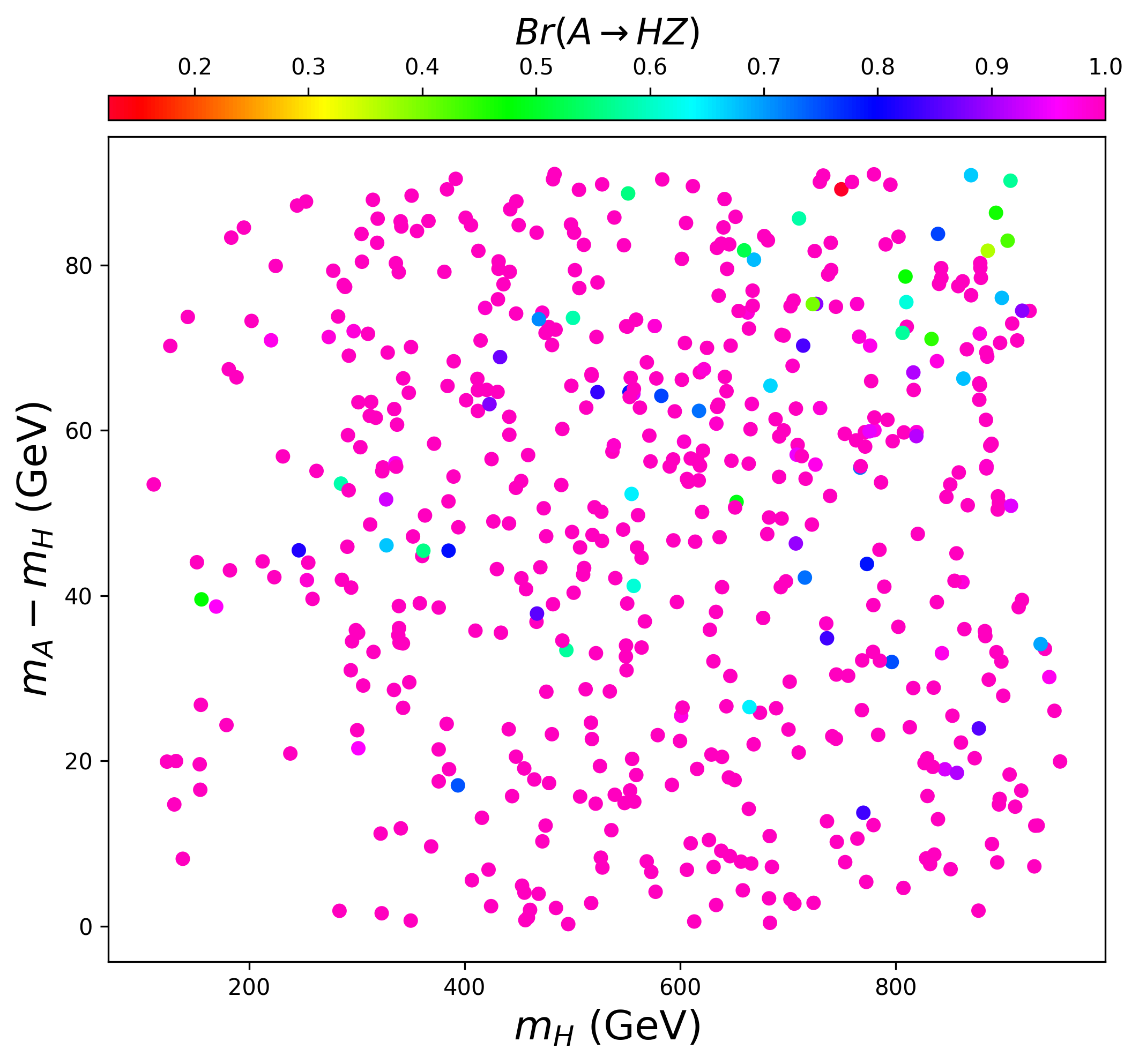}
	\includegraphics[width=0.4\textwidth]{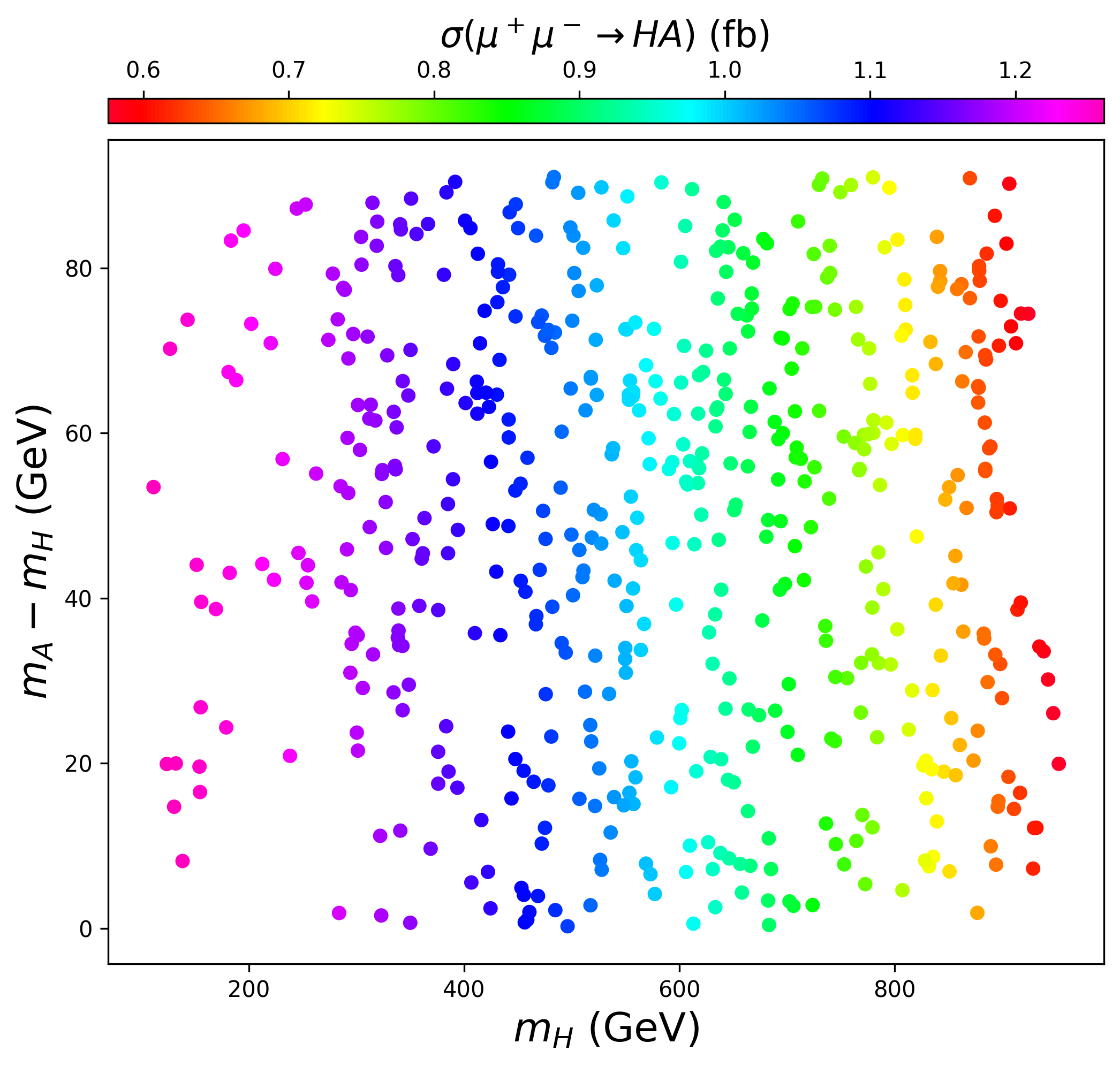}
	\caption{Viable parameter points in the $(m_A,\Delta m_{AH})$ plane (upper panels) and the $(m_H,\Delta m_{AH})$ plane (lower panels), where $\Delta m_{AH}=m_A-m_H$. The points are colour-coded according to $\operatorname{Br}(A\to HZ)$ in the left panels and the leading-order cross section for $\mu^+\mu^-\to HA$ at $\sqrt{s}=3~\mathrm{TeV}$ in the right panels.} 
	\label{fig:fig3}
\end{figure}

At the production level, the associated $HA$ pair is generated through the exchange of a virtual $Z$ boson in the $s$-channel, as illustrated in figure~\ref {fig:fig1}.
Figure~\ref{fig:fig2} displays the viable scan points in the $(m_H,m_A)$ plane, colour-coded according to the leading-order cross section for $\mu^+\mu^-\to HA$ at $\sqrt{s}=3~\mathrm{TeV}$. The points populate a narrow band close to the diagonal, reflecting the compressed-spectrum condition imposed in eq.~\ref{eq:compressed}. The colour gradient shows that the cross section decreases smoothly from approximately $1.2fb$ in the low-mass region to about $0.6fb$ for the heaviest allowed scalar states. Since the electroweak gauge structure entirely determines the $ZHA$ coupling, the tree-level production cross section at a fixed centre-of-mass energy depend directly on $m_H$ and $m_A$, while it is independent of the scalar potential parameters $\lambda_L$ and $\lambda_2$, as well as $m_{H^{\pm}}$.
Figure~\ref{fig:fig3} show the allowed points in the $(m_A, \Delta m_{AH})$ and ($m_H, \Delta m_{AH}$) planes.  The upper panels show the mass splitting $\Delta m_{AH}$ as a function of $m_A$, whereas the lower panels display the same quantity as a function of $m_H$. Both show similar distributions since $m_A\simeq m_H$.

The right panels show the cross section represented by the colour bar. It decreases from about $1.2$ to $0.6~\mathrm{fb}$ as the scalar masses increase. No clear dependence on $\Delta m_{AH}$ is observed over the selected range. he production cross section shows no clear dependence on $\Delta m_{AH}$ within the selected range. Its decrease is mainly due to the reduced phase space at higher scalar masses. However, $\Delta m_{AH}$ directly affects the phase space and fermion-pair kinematics of the three-body decay $A\to Hf\bar{f}$.

The left panels show $\operatorname{Br}(A\to HZ)$ through the colour bar. All selected points satisfy $\Delta m_{AH}<m_Z$, so the $Z$ boson is off-shell. The decay proceeds through $A\to HZ^*\to Hf\bar{f}$. Its branching ratio is close to unity for most points, with no clear dependence on $\Delta m_{AH}$. When kinematically allowed, the competing decay $A\to H^\pm W^{\mp}$ can lower this branching ratio, depending on the charged scalar mass $m_{H^\pm}$.

%
\subsection{VBF-enhanced}\label{subsec:vbf_production}
\begin{figure}[H]
	\centering
	\includegraphics[width=0.68\textwidth]{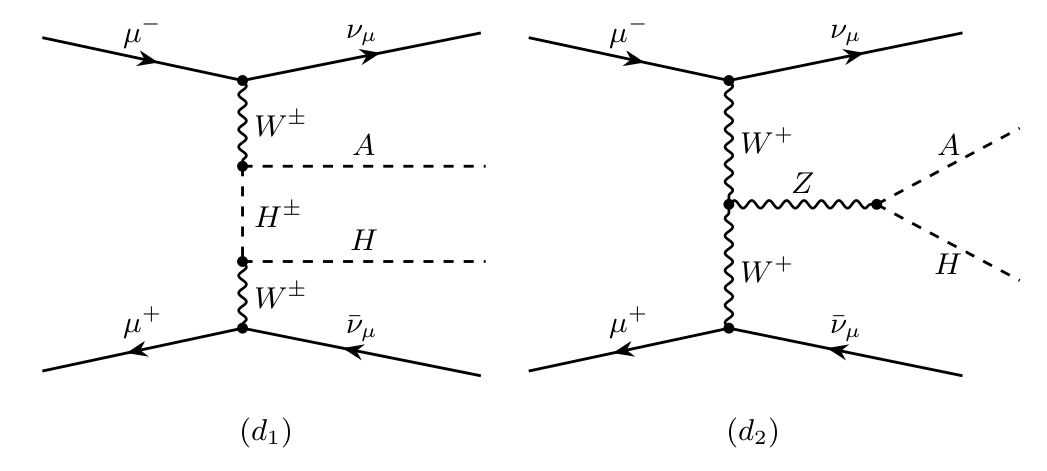}
	\caption{Main tree-level Feynman topologies contributing to the process $\mu^+\mu^-\to HA\nu_\mu\bar{\nu}_\mu$.}
	\label{fig:fig4}
\end{figure}

\begin{figure}[H]
	\centering
	\includegraphics[width=0.47\textwidth]{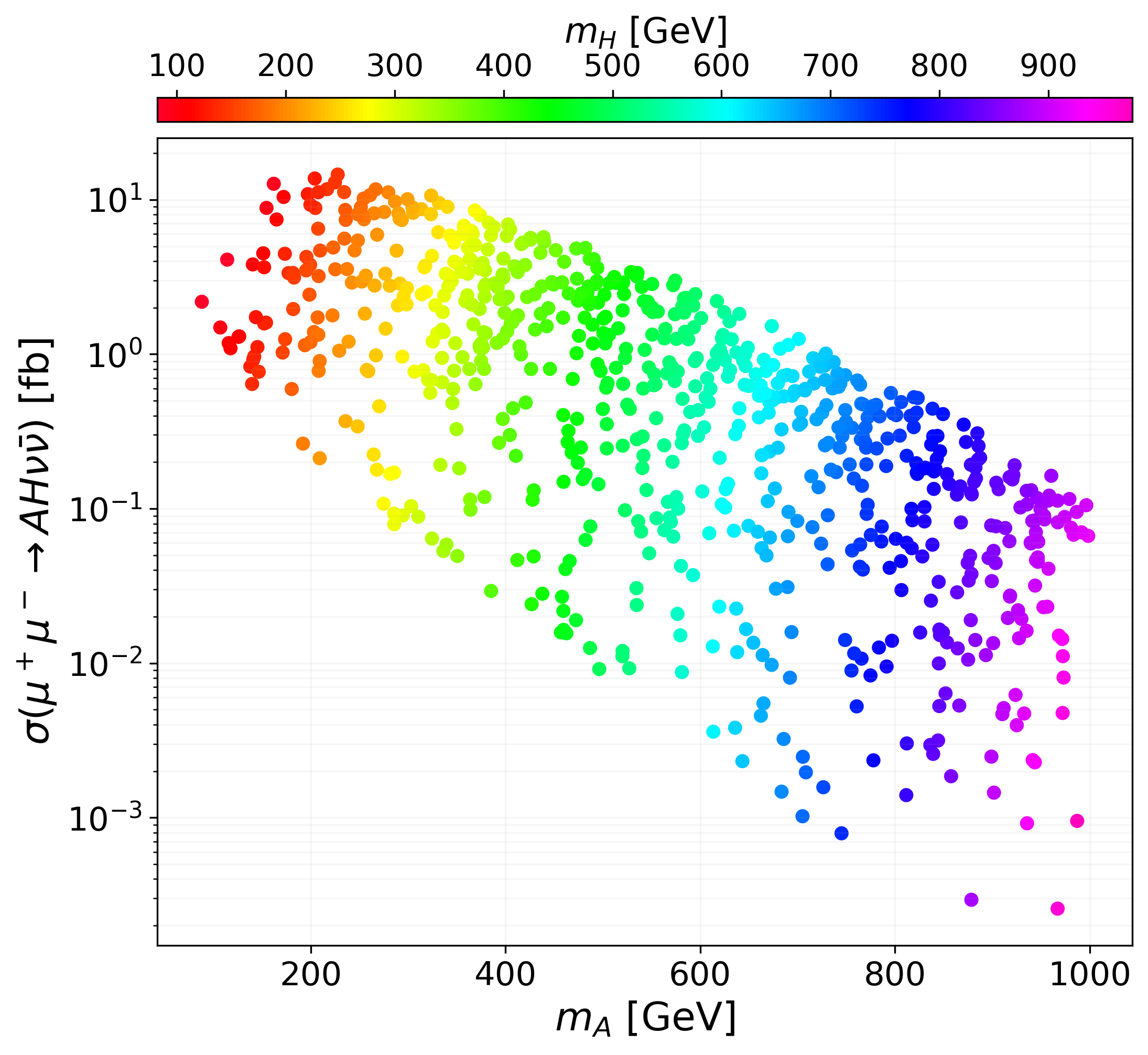}
	\includegraphics[width=0.47\textwidth]{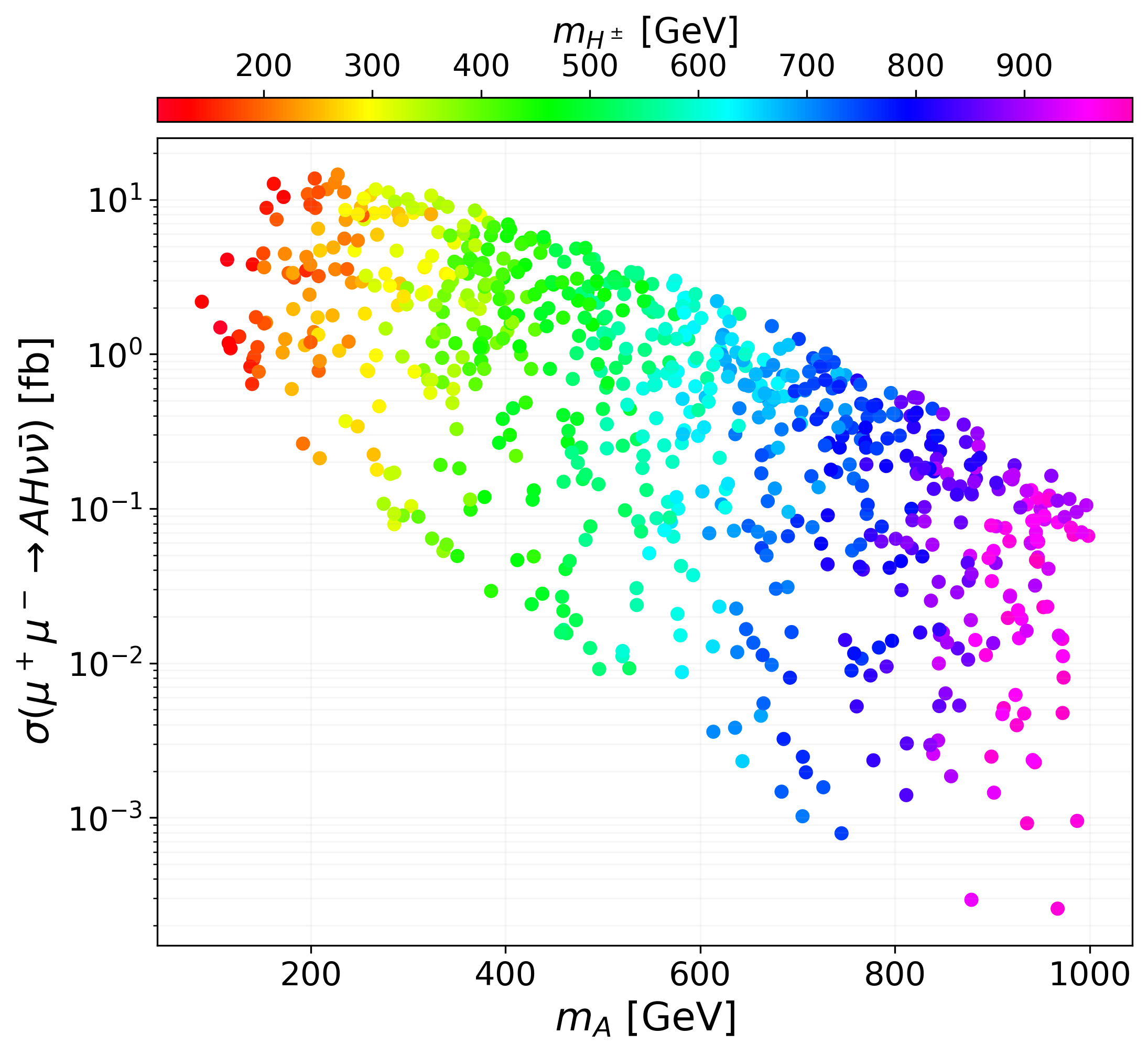}
	\caption{Leading-order cross section for $\mu^+\mu^-\to HA\nu_\mu\bar\nu_\mu$ at $\sqrt{s}=3$ TeV , shown as a function of $m_A$. 
    The colour scales denote $m_H$ (left) and $m_{H^{\pm}}$ (right).}
	\label{fig:fig5}
\end{figure}

The complementary production mode is
\begin{equation}
 \mu^+\mu^-\to HA\nu_\mu\bar\nu_\mu,
 \label{eq:vbf-production}
\end{equation}
which contains $W^+W^-$-fusion topologies. We refer to this channel as VBF enhanced: the numerical cross section is understood to be obtained from the complete gauge-invariant matrix element, rather than from an isolated diagram subset or an effective-vector-boson approximation. At the subprocess level, $W^+W^-\to HA$ receives coherent contributions from $s$-channel $Z$ exchange and charged-inert-scalar exchange. Representative contributions are shown in figure~\ref{fig:fig4}.
Figure~\ref{fig:fig5} we show the LO cross-section at $\sqrt{s}=3$TeV as a function of $m_A$, with the points colored by  $m_H$ and $m_{H^{\pm}}$. As expected at a fixed collider energy, the rate generally decreases as the invariant-mass threshold of the $HA$ system increases. The sizeable vertical spread at a given $m_A$ cannot be attributed to $m_H$ alone, because $m_H$ is already tightly correlated with $m_A$ by the compressed selection. It also reflects the unshown $m_{H^\pm}$ dependence of the charged-scalar propagators and the constructive or destructive interference among the gauge-related amplitudes. Raising the collider energy produces a pronounced enhancement characteristic of electroweak-boson fusion. This enhancement arises primarily from the increasing effective $W$ luminosity radiated by the incoming muons---often described as the growing $W$ content of the muon---rather than from an unrestricted growth of the hard $W^+W^-\to HA$ amplitude, whose high-energy behaviour is controlled by gauge cancellations.

\subsection{Benchmark scenarios}\label{subsec:benchmark}
For reproducibility, the IDM parameters and the principal dark-matter and collider observables defining the six benchmark points are summarized in Table~\ref{tab:benchmarks}. BP1 is selected as the reference scenario for the production cross-section study and the subsequent cut-based collider analysis. The remaining benchmark points, BP2–BP6, are considered in the machine-learning analysis to investigate the impact of different mass splittings, \(\Delta m=m_A-m_H\), on the signal discrimination performance. 
\begin{table}[H]
\centering
\renewcommand{\arraystretch}{1.12}
\setlength{\tabcolsep}{6pt}

\resizebox{\textwidth}{!}{%
\begin{tabular}{lcccccc}
\toprule
Parameters & BP1 & BP2 & BP3 & BP4 & BP5 & BP6 \\
\midrule
$m_H$ [GeV] 
& 314.8 & 311.91 & 311.43 & 317.5 & 315.35 & 310.25 \\

$m_A$ [GeV] 
& 402.69 & 358.73 & 335 & 379.01 & 348.53 & 381.92 \\

$m_{H^\pm}$ [GeV] 
& 451.32 & 316.58 & 396.4 & 416.64 & 425.15 & 438.91 \\
$\Delta m = m_A-m_H$ [GeV]
& 87.89 & 46.82 & 23.57 & 61.51 & 33.18 & 71.67 \\


$\lambda_2$ 
& 1 & 2.016 & 1.506 & 3.709 & 0.179 & 0.322 \\

$\lambda_L$ 
& 0.02 & 0.0332 & -0.0791 & -0.0667 & 0.0684 & 0.0834 \\
\midrule
$\Omega_H h^2$ 
& 0.000424 & 0.00613 & 0.00138 & 0.00846 & 0.000621 & 0.000429 \\

$\xi\sigma_{\rm SI}$ [pb] 
& $1.5\times10^{-10}$ & $3.96\times10^{-10}$ 
& $2.26\times10^{-9}$ & $1.54\times10^{-9}$ 
& $1.64\times10^{-9}$ & $2.52\times10^{-9}$ \\
\midrule
$\mathrm{BR}(A\to HZ)$ 
& 1 & 0.57 & 1 & 1 & 1 & 1 \\

$\sigma(\mu^+\mu^-\to HA)$ [fb] 
& 1.166 & 1.18 & 1.18 & 1.17 & 1.18 & 1.17 \\

$\sigma(\mu^+\mu^-\to HA\nu_\mu\bar\nu_\mu)$ [fb] 
& 6.92 & 2.2 & 0.658 & 3.49 & 1.17 & 4.76 \\

\bottomrule
\end{tabular}%
}

\caption{Definition of the six benchmark points used in the
collider analysis. The two production cross sections are
quoted at $\sqrt{s}=3$ TeV before decays and selection cuts.}
\label{tab:benchmarks}

\end{table}
\section{Cut-based analysis}\label{sec:cutbased}
In the previous section, we evaluated the parton-level cross sections for the proposed channels aimed at probing neutral inert scalars in the compressed regime of the IDM. Although these cross sections are not negligible, the actual discovery potential strongly depends on our ability to efficiently discriminate the signal from the overwhelming backgrounds. Our approach incorporates hard-scattering matrix elements, resonance decays, parton showering, hadronization, hadron decays, and a simplified detector response. 

Focusing on a muon collider with $\sqrt{s}=3$ TeV, we perform a detailed analysis of the two processes $\mu^+\mu^- \to HA$ and $\mu^+\mu^- \to HA\nu_\mu\bar\nu_\mu$, followed by $A \to H Z^* \to H \ell^+ \ell^-$, with $\ell=e,\mu$. Signal and background events are generated within a common Monte Carlo framework. The input parameters in the \texttt{MadGraph5} \cite{Alwall:2014hca} cards are adjusted according to the slected IDM benchmarks. The generation-level lepton acceptance requirements are $p_T(\ell)>5~\mathrm{GeV}$, $|\eta(\ell)|<2.5$, and $\Delta R(\ell,\ell)\geq0.4$, where $\Delta R=\sqrt{(\Delta\eta)^2+(\Delta\phi)^2}$. The generated parton-level events are passed to \texttt{Pythia~8.243} \cite{Sjostrand:2007gs} for parton showering, hadronization, and hadron decays, followed by a fast detector simulation using \texttt{Delphes~3.4.2} \cite{deFavereau:2013fsa} with the MuCs Detector TARGET model. Jets are reconstructed using the anti-$k_T$ algorithm \cite{Cacciari:2008gp} with a radius parameter R=0.4. Pileup and beams-induced background specific to muon colliders are not explicitly simulated, and multiple parton interactions from soft QCD processes are switched off at the \texttt{Pythia} level. The resulting sensitivities are therefore interpreted  as idealised assessments under the adopted detector and event-selection assumptions. The simulated samples befor the channel-specific selections are referred to below as initial events. Finally, both signals and background events are analyzed using \texttt{MadAnalysis5} \cite{Conte:2013mea}.
\subsection{Signal signatures and background processes}\label{subsec:analysis-setup}

The two signal categories are therefore
\begin{align}
 \mu^+\mu^-&\to HA\to HH\ell^+\ell^- ,
 \label{eq:direct-analysis-chain}\\
 \mu^+\mu^-&\to HA\nu_\mu\bar\nu_\mu
 \to HH\ell^+\ell^-\nu_\mu\bar\nu_\mu .
 \label{eq:vbf-analysis-chain}
\end{align}

\begin{figure}[H]
	\centering\includegraphics[width=0.52\textwidth]{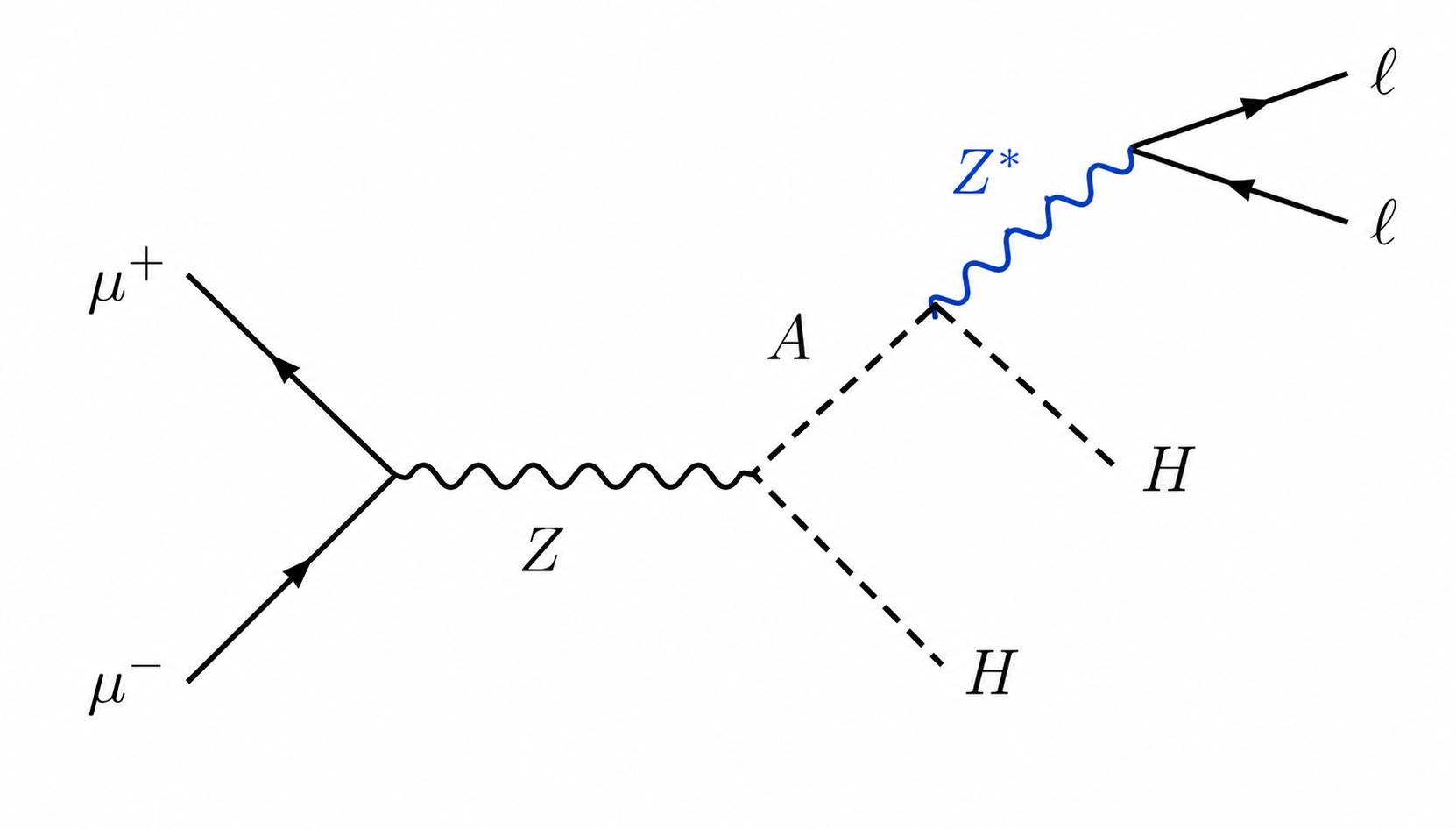}
	\caption{Representative tree-level diagram for the direct signal $\mu^+\mu^-\to HA\to HHZ^*\to HH\ell^+\ell^-$.}
	\label{fig:fig6}
\end{figure}

\begin{figure}[H]
	\centering\includegraphics[width=0.72\textwidth]{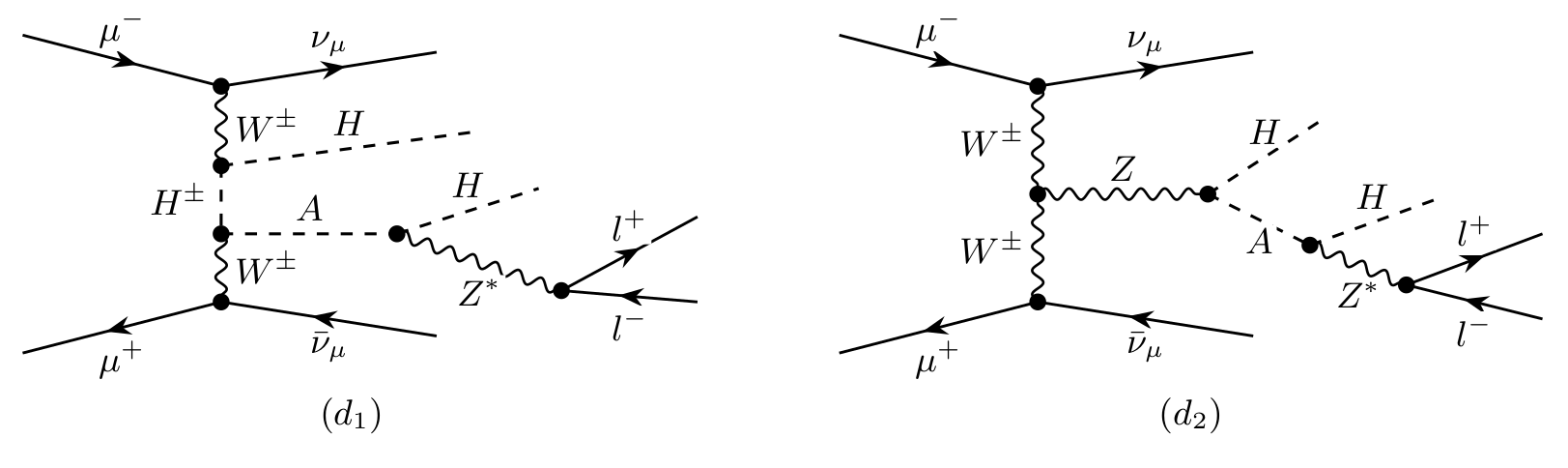}
	\caption{Representative diagrams for the VBF-enhanced signal $\mu^+\mu^-\to HA\nu_\mu\bar\nu_\mu$, followed by $A\to HZ^*\to H\ell^+\ell^-$.}
	\label{fig:fig7}
\end{figure}

\begin{figure}[H]
	\centering
	\includegraphics[width=0.45\textwidth]{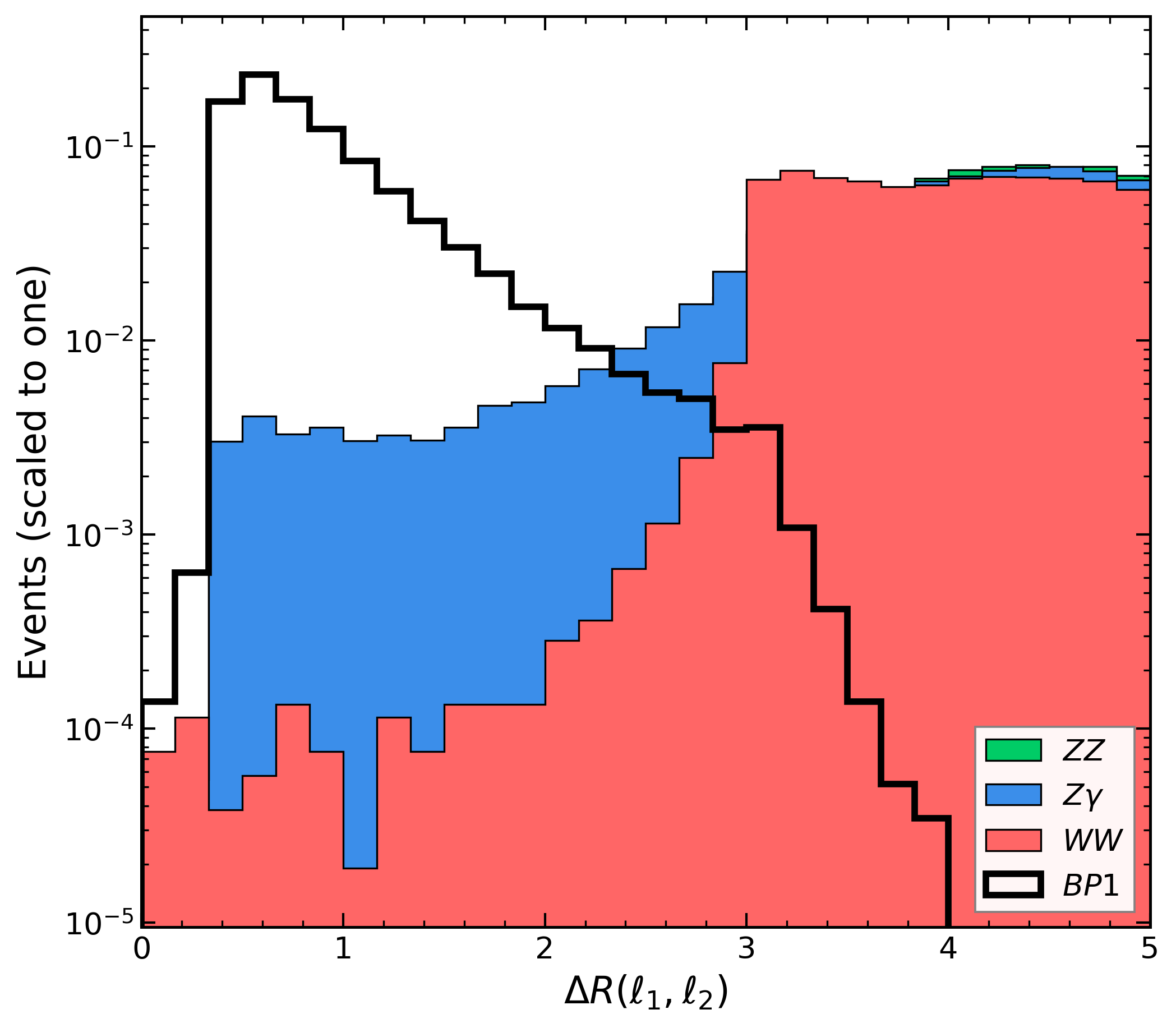}
	\includegraphics[width=0.45\textwidth]{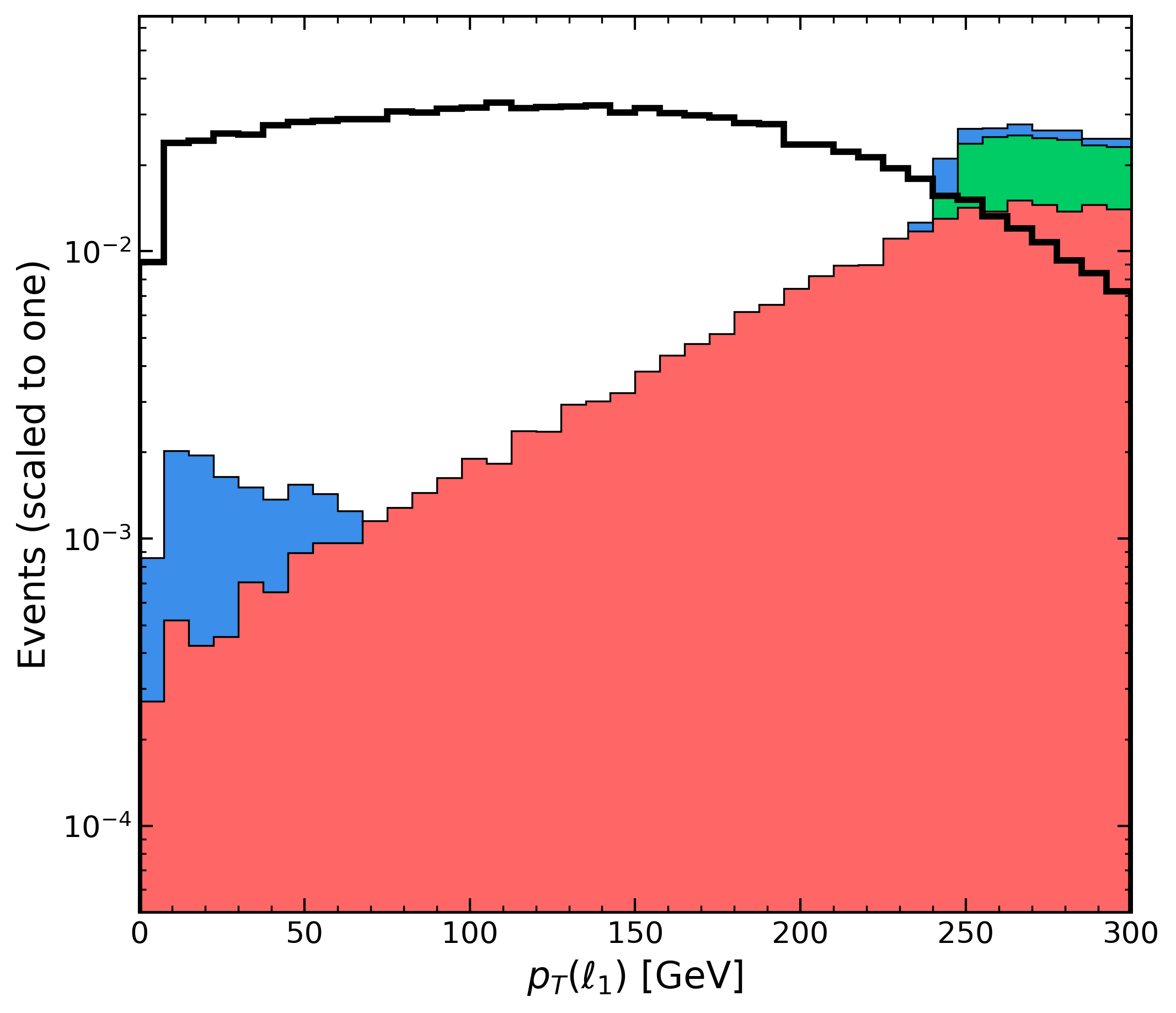}
	\includegraphics[width=0.45\textwidth]{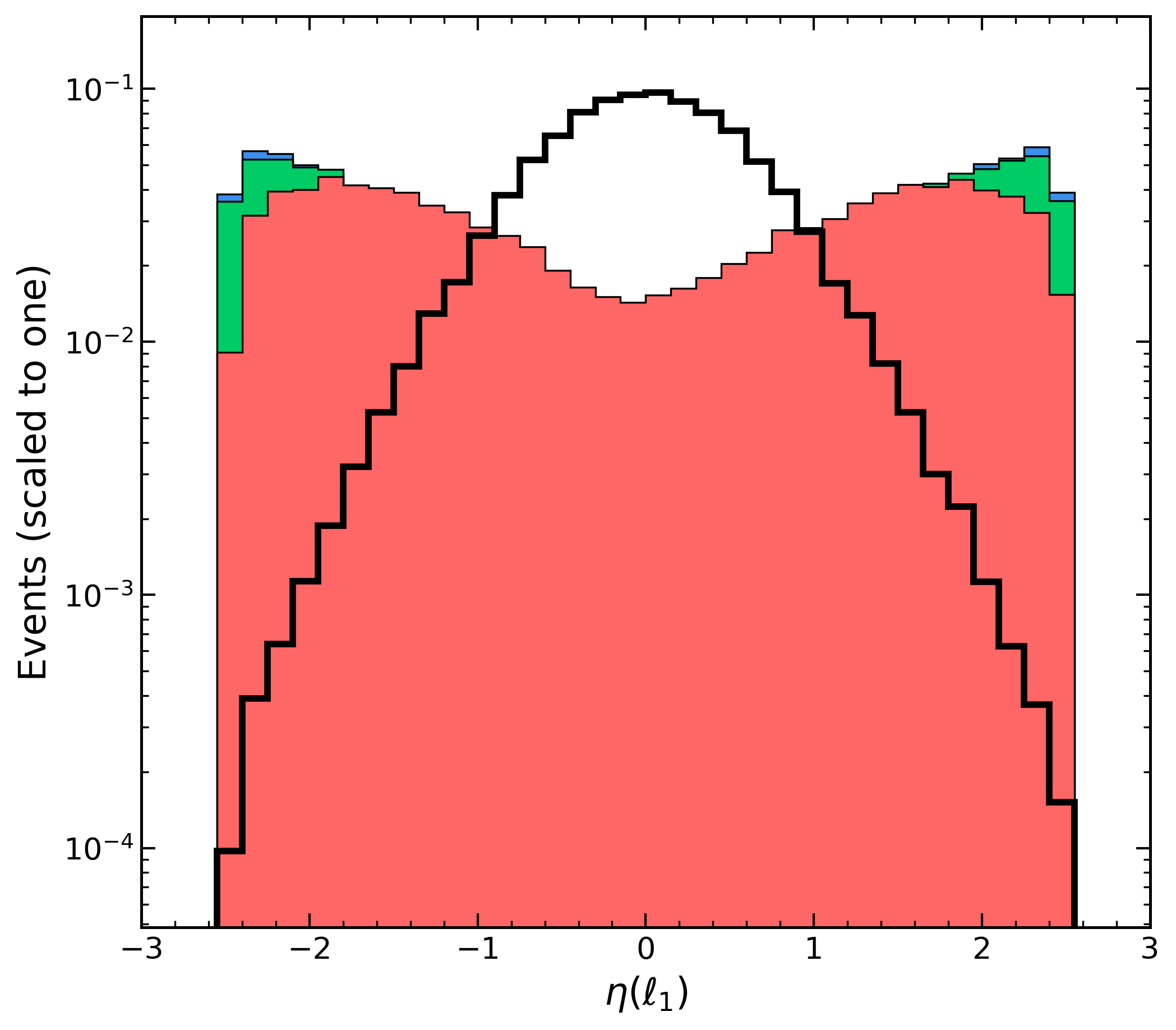}
	\includegraphics[width=0.45\textwidth]{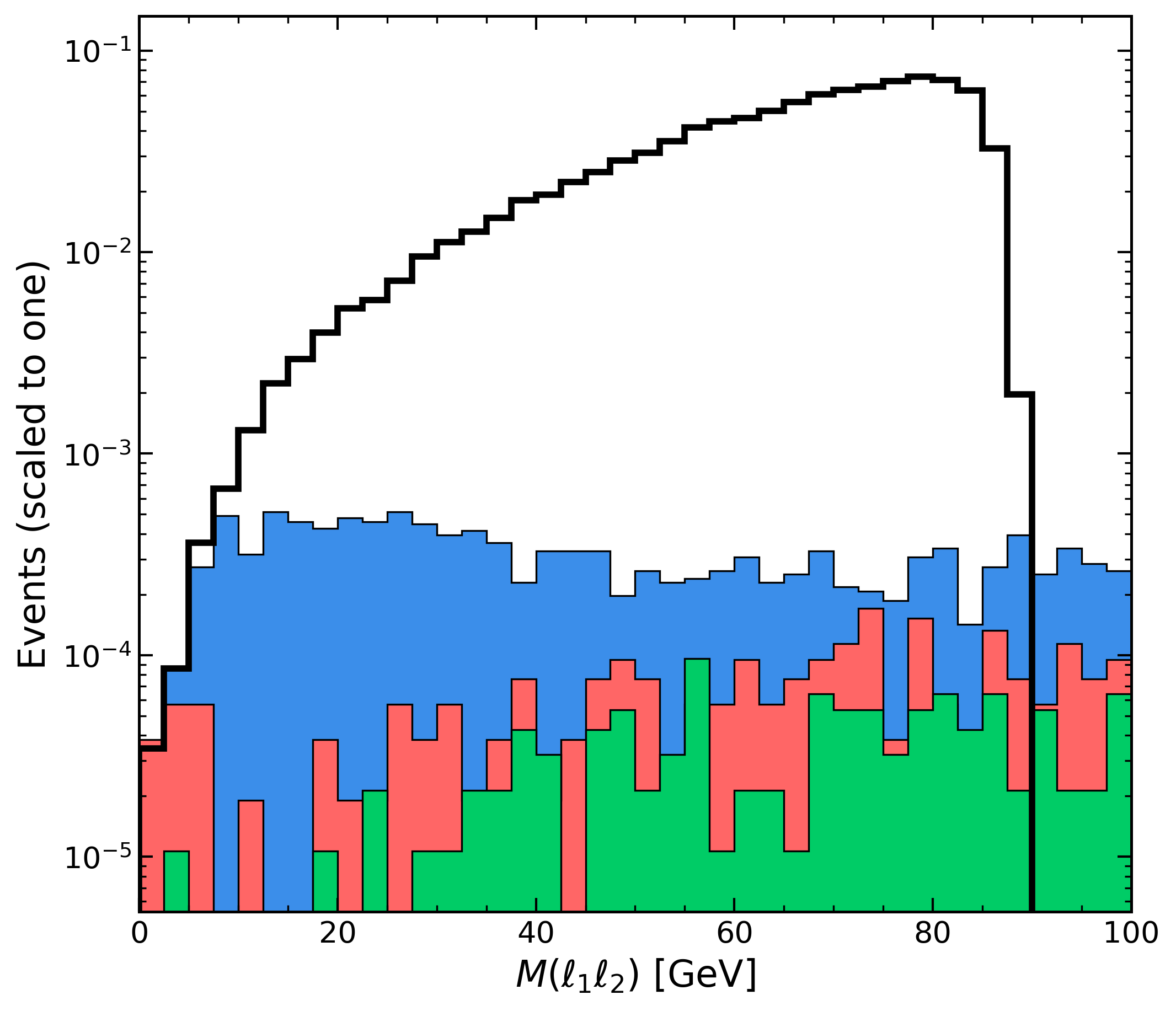}
	\caption{Normalised kinematic distributions for the direct-production final state $HH\ell^+\ell^-$ at a $\sqrt{s}=3$ TeV muon collider: $\Delta R(\ell_1,\ell_2)$, $p_T(\ell_1)$, $\eta(\ell_1)$, and $m_{\ell\ell}$.} 
	\label{fig:analy1}
\end{figure}

\begin{figure}[H]
	\centering
	\includegraphics[width=0.45\textwidth]{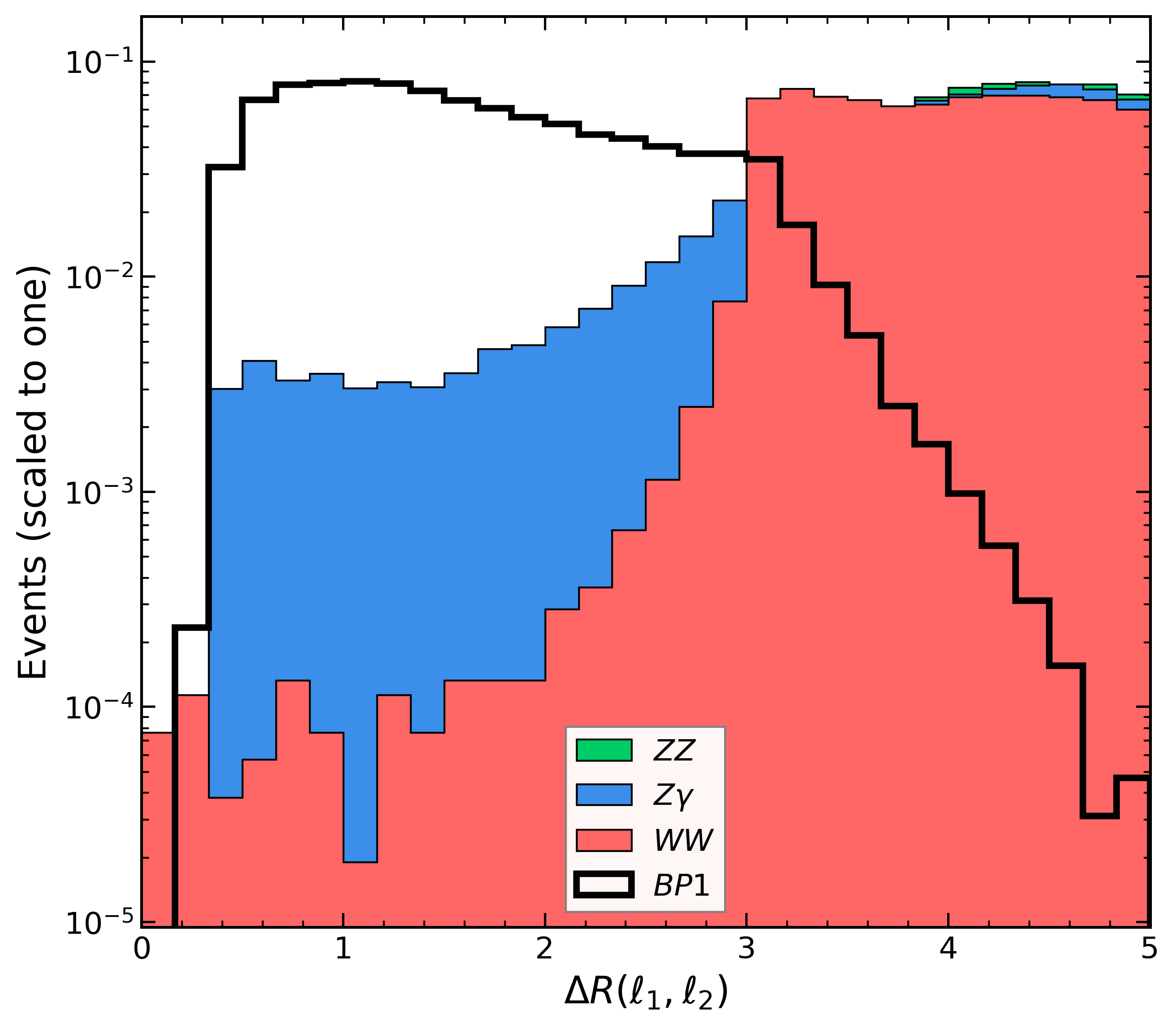}
	\includegraphics[width=0.45\textwidth]{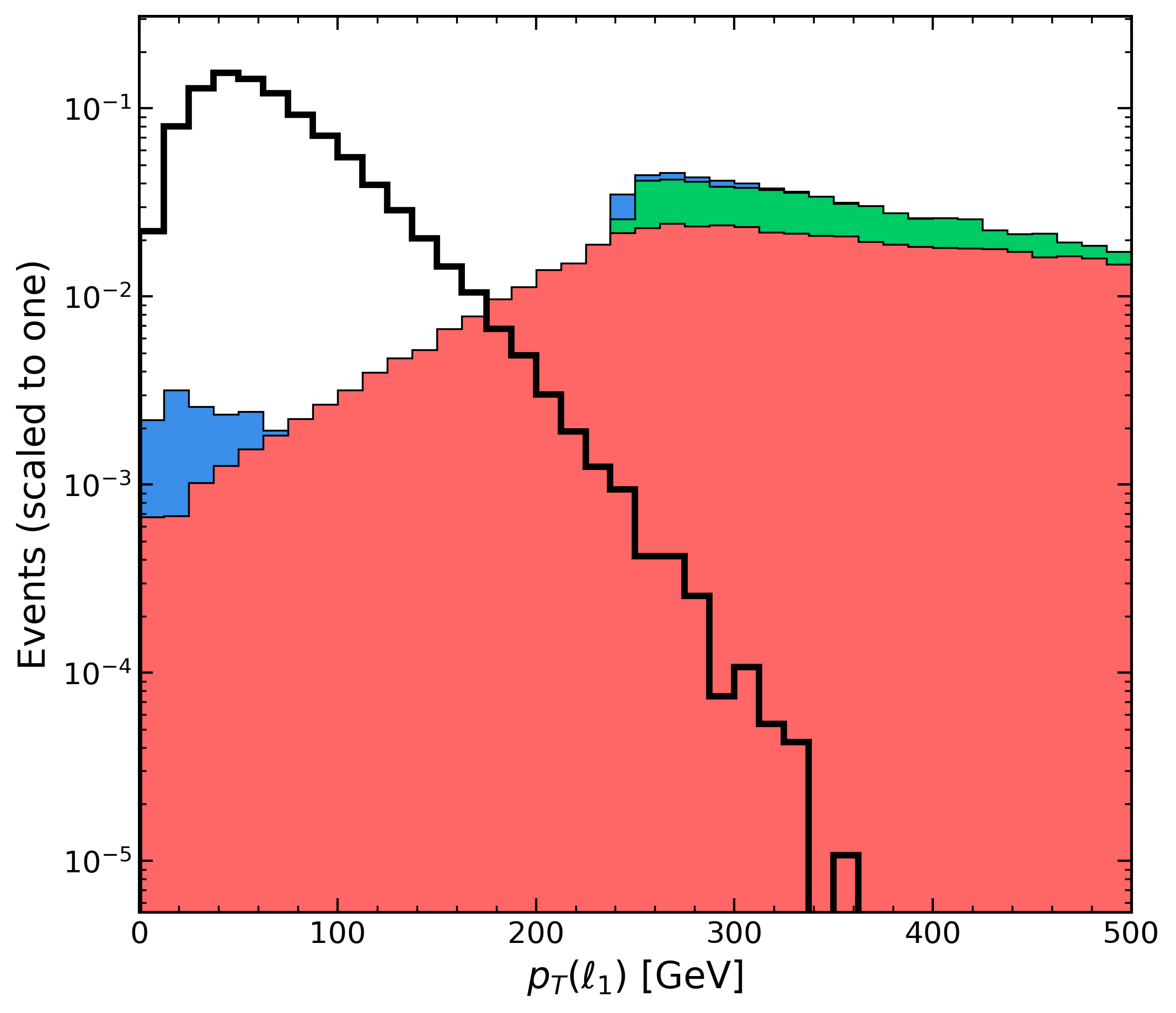}
	\includegraphics[width=0.45\textwidth]{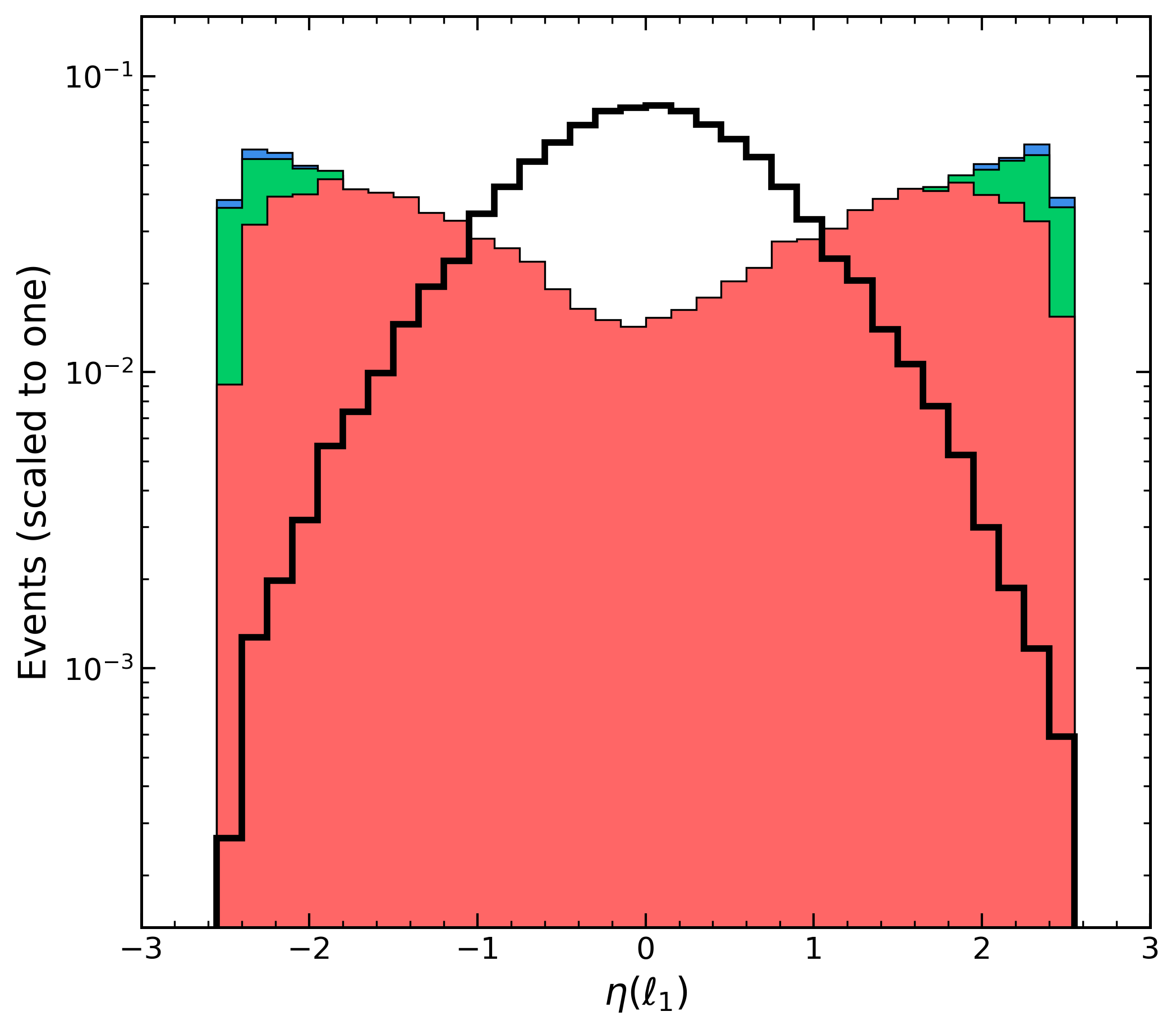}
	\includegraphics[width=0.45\textwidth]{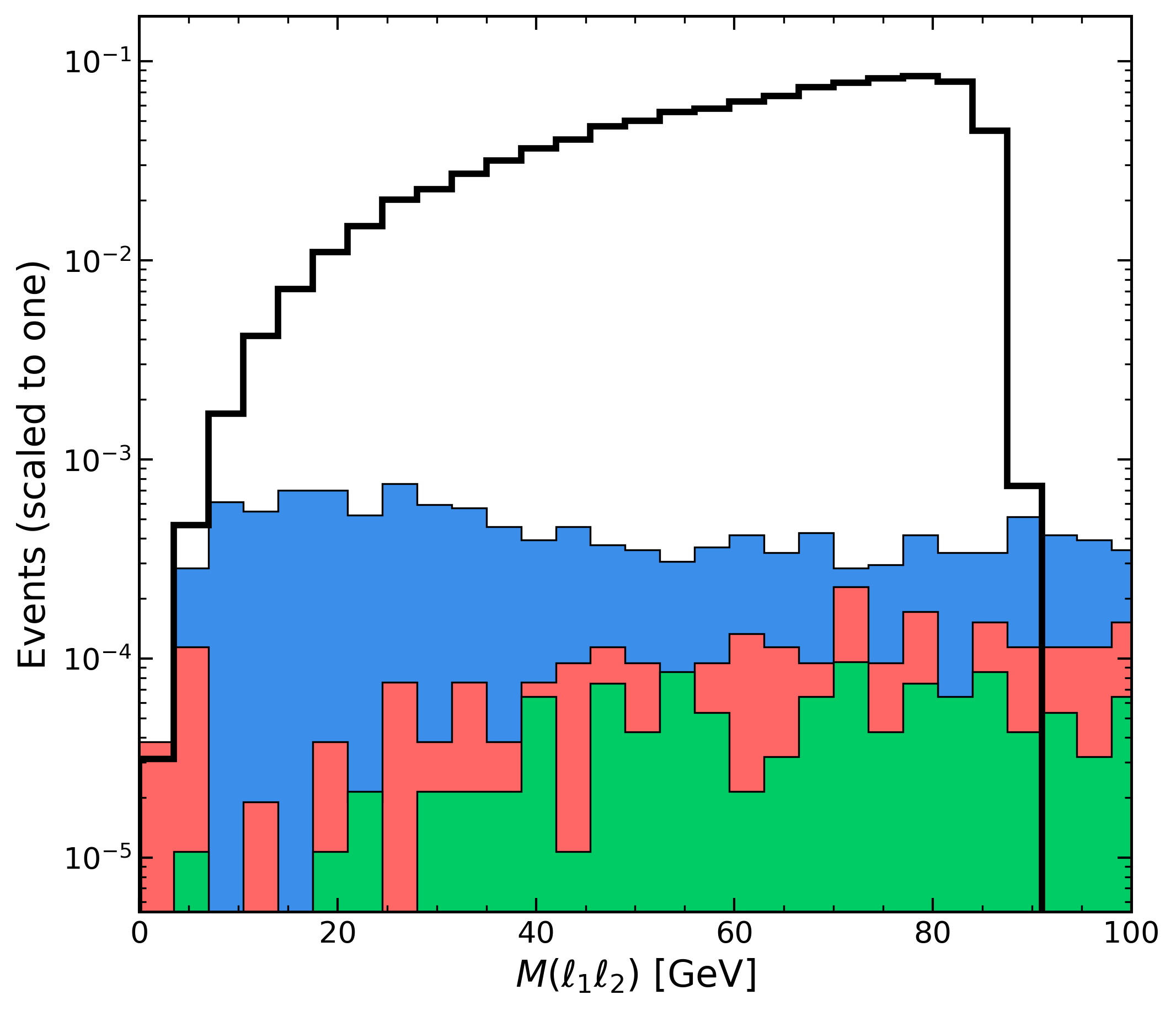}
	\caption{Normalised distributions of $\Delta R(\ell_1,\ell_2)$, $p_T(\ell_1)$, $\eta(\ell_1)$, and $m_{\ell\ell}$ for the $HH\ell^+\ell^-\nu_\mu\bar\nu_\mu$ final state at a muon collider with $\sqrt{s}=3~\mathrm{TeV}$.}
	\label{fig:analy2}
\end{figure}

Their representative diagrams are shown in Figures~\ref{fig:fig6} and~\ref{fig:fig7}, respectively. Both signatures contain an opposite-sign, same-flavour dilepton pair accompanied by missing momentum from the stable inert scalars. The VBF-enhanced channel additionally contain two neutrinos. As in the production-level study, this channel is evaluated using the complete matrix element, including all contributing diagrams and their interference.
\begin{samepage}
We use a common set of Standard Model backgrounds so that the direct and VBF-enhanced selections can be compared consistently:
\begin{enumerate}[label=(\roman*)]
    \item $\mu^+\mu^- \to W^+W^- \to \ell^+\nu\,\ell^-\bar{\nu}$ (WW).
    \item $\mu^+\mu^- \to ZZ \to \ell^+\ell^-\,\nu\bar{\nu}$ (ZZ). 
    \item $\mu^+\mu^- \to Z^*/\gamma^* \to \ell^+\ell^- jj$ (DY).
\end{enumerate}
\end{samepage}
The two leptons are ordered in transverse momentum, $p_T(\ell_1)>p_T(\ell_2)$. Figures~\ref{fig:analy1} and~\ref{fig:analy2} compare the distributions of $\Delta R_{\ell\ell}, p_T(\ell_1), \eta_T(\ell_1)$, and $m_{\ell\ell}$ for the signal and backgrounds. These normalised distributions illustrate differences in shape; the relative event rates are given by the cut-flow tables.
For the direct-production channel, Figure~\ref{fig:analy1} show the distribution for BP1. The signal is concentered at relatively small dilepton angular separation, whereas the backgrounds have substantial contributions at larger $\Delta R_{\ell\ell}$. This behaviour motivates the requirement $\Delta R_{\ell\ell} < 2.3$. According to Table~\ref{tab:cutflow:3TeV_s_chan}, this selection reduces the signal cross section from $0.057$ to $0.0322~\mathrm{fb}$, retaining approximately $56.5\%$ of the events passing the basic cuts. At the same time the total background cross section decreases from $11.608$ to $0.1509~\mathrm{fb}$ with the strongest suppression observed in the $WW$ channel, which decreases from $6.168$ to $0.0044~\mathrm{fb}$. 

The pseudorapidity distribution in the lower-left panel of Figure~\ref{fig:analy1} exhibits a central signal peak, while the background distributions have larger contributions away from $\eta=0$. We therefore impose $|\eta(\ell_1)|<0.9$. This requirement retains approximately $87.0\%$ of the signal surviving the angular selection and reduces the total background to $0.0547~\mathrm{fb}$. The leading-lepton transverse-momentum distribution provides further discrimination: the signal is broadly distributed over lower and intermediate $p_T$ values, while the backgrounds have a harder component in the displayed range. 

The cut $p_T(\ell_1)<240~\mathrm{GeV}$ reduces the total background to $5.93\times10^{-3}~\mathrm{fb}$. It retains $80.0\%$ of the signal from the previous selection. No $ZZ$ events survive this cut in the simulated sample. The lower-right panel of Figure~\ref{fig:analy1} shows the dilepton invariant-mass distribution. Neglecting the lepton masses, the three-body decay $A\to H\ell^+\ell^-$ has the parton-level endpoint
\begin{equation}
m_{\ell\ell}^{\max}=m_A-m_H.
\end{equation}
For BP1 the masses listed in Table~\ref{tab:benchmarks} give $m_A-m_H=87.89~\mathrm{GeV}$. This agrees with the sharp drop observed near the upper edge of the signal distribution. The spectrum increases towards the endpoint and then falls rapidly. We therefore impose $m_{\ell\ell}<87~\mathrm{GeV}$, close to the kinematic limit. This cut has little effect on the signal cross section, which decreases from $0.0224$ to $0.0223~\mathrm{fb}$. The remaining $WW$ and $Z^*/\gamma^*$ backgrounds are reduced to $2.4\times10^{-4}$ and $4.1\times10^{-3}~\mathrm{fb}$, respectively.

After the full selection, about $39.1\%$ of the signal remains relative to the basic cuts. The total background cross section drops to $4.34\times10^{-3}~\mathrm{fb}$. The signal-to-background ratio increases from $4.91\times10^{-3}$ to about $5.14$. The remaining background is mainly due to the $Z^*/\gamma^*$ process.

The angular and centrality cuts suppress most of the background at the initial stages. The upper cut on the leading-lepton transverse momentum improves the signal purity. The final invariant-mass cut removes more background with only a small loss of signal.

\begin{table}[H]
\centering
{\footnotesize\renewcommand{\arraystretch}{1.05}
\setlength\tabcolsep{9pt}
\begin{tabular}{lcccc}
\toprule
Selection & $WW$ & $ZZ$ & $Z^*/\gamma^*$ & Signal \\
& \multicolumn{4}{c}{Cross section [fb]}\\
\midrule
Basic cuts&  6.168 & 4.05 & 1.39  & 0.057\\
$\Delta R(\ell_1,\ell_2)<2.3$ & 0.0044 & 0.0875 & 0.059 & 0.0322\\
$|\eta(\ell_1)|<0.9$ & 0.0027 & 0.028 & 0.024& 0.028\\
$p_T(\ell_1)<240~\gev$ &0.00043 & 0 & 0.0055 & 0.0224\\
$m_{\ell\ell}<87~\gev$ &0.00024 & 0 & 0.0041 & 0.0223\\
\bottomrule
\end{tabular}
}
\caption{Cut flow for $\mu^+\mu^-\to HA\to HH\ell^+\ell^-$ at $\sqrt{s}=3$ TeV.}
\label{tab:cutflow:3TeV_s_chan}
\end{table}
We next consider the VBF-enhanced signal, whose kinematic distributions are shown in Figure.~\ref{fig:analy2}. The signal again favours smaller dilepton angular separations than the principal background contributions, but its distribution extends over a broader interval than in Figure.~\ref{fig:analy1}. The selection in Table~\ref{tab:cutflow:3TeV_vbf} uses $0.2<\Delta R(\ell_1,\ell_2)<2.9$. This requirement retains approximately $58.2\%$ of the initial signal, reducing its cross section from $0.414$ to $0.241~\mathrm{fb}$. The total background drops from $11.608$ to $0.3504~\mathrm{fb}$. The $WW$ contribution falls to $0.0254~\mathrm{fb}$.

The signal leading lepton is mainly produced in the central region, as shown in the lower-left panel of Figure.~\ref{fig:analy2}. The cut $|\eta(\ell_1)|<1$ reduces the background cross section to $0.1555~\mathrm{fb}$. It retains about $83.4\%$ of the signal surviving the angular cut.

The upper-right panel shows a clear separation in the leading-lepton transverse momentum. The signal peaks at low $p_T$ and falls rapidly at higher values. In contrast, the backgrounds extend to larger transverse momenta. We therefore impose $p_T(\ell_1)<170~\mathrm{GeV}$. This cut reduces the signal cross section from $0.201$ to $0.193~\mathrm{fb}$, retaining about $96.0\%$ of the signal. The total background decreases to $0.0100~\mathrm{fb}$.
The dilepton $m_{\ell\ell}$ distribution shows the endpoint from $A\to H\ell^+\ell^-$. The additional neutrinos contribute to the missing transverse momentum and affect the recoil of the $HA$ system. However, they do not change the dilepton invariant-mass endpoint, which remains fixed by $m_A-m_H$. However, the parton-level upper bound on $m_{\ell\ell}$ is still determined by $m_A-m_H$. After applying the final cut $m_{\ell\ell}<87~\mathrm{GeV}$, the signal cross section remains $0.193~\mathrm{fb}$, within the precision reported in Table~\ref{tab:cutflow:3TeV_vbf}. The surviving $WW$ and $Z^*/\gamma^*$ backgrounds are $2.41\times10^{-3}$ and $4.60\times10^{-3}~\mathrm{fb}$, respectively, while the tabulated $ZZ$ contribution is zero.

The full VBF-enhanced selection therefore retains approximately $46.6\%$ of the signal passing the basic cuts. The total quoted background is $7.01\times10^{-3}~\mathrm{fb}$, giving $S/B\simeq27.5$. For the samples and normalisations reported in Tables~\ref{tab:cutflow:3TeV_s_chan} and~\ref{tab:cutflow:3TeV_vbf}, the VBF-enhanced selection yields a larger surviving signal rate and a larger signal-to-background ratio.
\begin{table}[H]
\centering
{\footnotesize\renewcommand{\arraystretch}{1.05}
\setlength\tabcolsep{9pt}
\begin{tabular}{lcccc}
\toprule
Selection & $WW$ & $ZZ$ & $Z^*/\gamma^*$ & Signal \\
& \multicolumn{4}{c}{Cross section [fb]}\\
\midrule
Basic cuts&  6.168 & 4.05 & 1.39  & 0.414\\
$0.2<\Delta R(\ell_1,\ell_2)<2.9$ & 0.0254 & 0.207 & 0.118 & 0.241\\
$|\eta(\ell_1)|<1$ & 0.0219 & 0.0811 & 0.0525 & 0.201\\
$p_T(\ell_1)<170~\gev$ &0.0043 & 0 & 0.0057 & 0.193\\
$m_{\ell\ell}<87~\gev$ &0.00241 & 0 & 0.0046 & 0.193\\
\bottomrule
\end{tabular}
}
\caption{Cut flow for $\mu^+\mu^-\to HA\nu_\mu\bar\nu_\mu\to HH\ell^+\ell^-\nu_\mu\bar\nu_\mu$ at $\sqrt{s}=3$ TeV.}
\label{tab:cutflow:3TeV_vbf}
\end{table}
\subsection{Statistical significance}\label{subsec:statistics}

To quantify the sensitivity of the considered signatures, we evaluate the statistical significance using
\begin{equation}
\mathcal{Z}=\sqrt{\mathcal{L}}\frac{\sigma_S}{\sqrt{\sigma_S+\sigma_B}},
\label{eq:significance}
\end{equation}
where $\sigma_S$ and $\sigma_B$ denote the signal and background cross sections after all selection cuts, respectively, while $\mathcal{L}$ is the integrated luminosity. Using the values shown in Tables~\ref{tab:cutflow:3TeV_s_chan} and~\ref{tab:cutflow:3TeV_vbf} gives the estimates in Table~\ref{tab:baseline-significance}.

\begin{table}[H]
\centering
{\footnotesize
\renewcommand{\arraystretch}{1.05}
\setlength{\tabcolsep}{8pt}
\begin{tabular}{lcccc}
\toprule
Channel & $500~\mathrm{fb}^{-1}$ & $1000~\mathrm{fb}^{-1}$ & $1500~\mathrm{fb}^{-1}$ & $10~\mathrm{ab}^{-1}$\\
\midrule
Direct $HH\ell^+\ell^-$ (BP1) & 3.06 & 4.32 & 5.29 & 13.66 \\
VBF-enhanced $HH\ell^+\ell^-\nu\bar\nu$ & 9.65 & 13.65 & 16.71 & 43.16 \\
\bottomrule
\end{tabular}
}
\caption{Baseline counting sensitivity $\mathcal{Z}$ for the direct and VBF-enhanced channels at $\sqrt{s}=3$ TeV.}
\label{tab:baseline-significance}
\end{table}

At $\mathcal{L}=1000~\mathrm{fb}^{-1}$, the counting significance reaches $\mathcal{Z}\simeq4.32$ for direct production and $\mathcal{Z}\simeq13.65$ for the VBF-enhanced channel.

These values provide an initial comparison between the two channels and should not be taken as final discovery significances. The Monte Carlo statistical uncertainty must therefore be included in the significance estimate.
The cut-based analysis also provides a basis for the multivariate study in Section~\ref{sec:mva}.

\section{Multivariate analysis}\label{sec:mva}

We use two boosted decision tree (BDT) classifiers based on \texttt{XGBoost}~\cite{Chen:2016btl}. The first separates the signal from the backgrounds. The second divides the events into two orthogonal categories: VBF-enriched and s-channel-enriched. The output of the first classifier is then used as a binned discriminant for the statistical analysis in each category.

Prior to training, events passing the pre-selection described in Sec.~\ref{sec:cutbased} are rotated in the transverse plane such that the leading lepton defines $\phi=0$. The azimuthal coordinates of the subleading lepton, the dilepton system, and the missing transverse momentum are transformed accordingly:
\begin{equation}
    \phi_a^{\mathrm{rot}}=
    \operatorname{atan2}\left[\sin(\phi_a-\phi_{\ell_1}),
    \cos(\phi_a-\phi_{\ell_1})\right].
    \label{eq:idm_rotation}
\end{equation}
This transformation preserves the relative angular configuration of the event while removing its arbitrary absolute azimuthal orientation. The leading-lepton rotated coordinate is retained in the input list, although it is identically zero and provides no independent discrimination. No standardization or normalization of the input features is applied, since the tree-based classifiers do not require variables to be expressed on a common scale.

The selected events are divided independently within each simulated sample into training, test subsets containing approximately 75\%, and 25\% of the events, respectively. The test subset is reserved for performance evaluation and statistical analysis. Each event carries the physical normalization weight
\begin{equation}
    w_i^{\mathrm{phys}}=
    \frac{w_i^{\mathrm{MC}}}{\sum_j w_j^{\mathrm{MC}}}
    \sigma\mathcal{L}\mathcal{B},
    \label{eq:idm_event_weight}
\end{equation}
where the sum runs over the generated sample, $\sigma$ denotes the sample cross-section, and $\mathcal{B}$ accounts for branching fractions not already included in that cross-section. Additional rescaling of these weights is applied during training to balance the classes, as described below. Expected yields are always evaluated using the physical normalization, with the training subset restored to the full expected yield of each sample.

\subsection{Signal-versus-background XGBoost classifier}
\label{sec:idm_signal_bdt}
The first model is a parameterized XGBoost classifier designed to identify the dilepton-plus-missing-momentum signal against the background samples. The signal training set includes both VBF and s-channel production modes. To account for the variation of the signal kinematics across the benchmark hypotheses, the model is parameterized in terms of $m_H$ and $\Delta m$:
\begin{equation}
    D_{\mathrm{sig}}=f(\mathbf{x};m_H,\Delta m),
    \label{eq:idm_parameterized_bdt}
\end{equation}
where $\mathbf{x}$ contains the reconstructed event observables. For signal events, the conditioning parameters are set to the generated benchmark values, while for the  background events are conditioned with the training hypotheses. At evaluation, all background events are assigned the parameters of the signal hypothesis being tested. This ensures that the parameter is not used as a discriminant during the training.

The classifier uses 17 kinematic inputs and two conditioning parameters, as summarized in Table~\ref{tab:idm_bdt_inputs}. The lepton transverse momenta, angular separation, and dilepton invariant mass provide information about the off-shell decay topology. The missing transverse momentum, recoil mass characterize the invisible system and $\Delta R_{\ell\ell}$.

\begin{table}[htb!]
    \centering
    \begin{tabular}{ll}
    \hline\hline
    Objects or system & Training variables \\
    \hline
    Leading lepton & $p_T^{\ell_1}$, $|\eta_{\ell_1}|$, $\phi_{\ell_1}^{\mathrm{rot}}$ \\
    Subleading lepton & $p_T^{\ell_2}$, $|\eta_{\ell_2}|$, $\phi_{\ell_2}^{\mathrm{rot}}$ \\
    Dilepton system & $m_{\ell\ell}$, $p_T^{\ell\ell}$, $\Delta R_{\ell\ell}$, $\phi_{\ell\ell}^{\mathrm{rot}}$ \\
    Missing transverse momentum & $p_T^{\mathrm{miss}}$, $\phi_{\mathrm{miss}}^{\mathrm{rot}}$ \\
    Event-level observables & $m_{\mathrm{recoil}}$, $\Delta\phi_{\ell\ell,\mathrm{miss}}$, $J_{\mathrm{bal}}$, $m_T^{(p)}$, $m_T^{(E)}$ \\
    Conditioning parameters & $m_H$, $\Delta m$ \\
    \hline\hline
    \end{tabular}
    \caption{Input variables used by the signal-versus-background classifier.}
    \label{tab:idm_bdt_inputs}
\end{table}

The recoil mass is calculated from the dilepton energy $E_{\ell\ell}$ and the nominal collision energy as
\begin{equation}
    m_{\mathrm{recoil}}^2=s+m_{\ell\ell}^2-2\sqrt{s}E_{\ell\ell}.
    \label{eq:idm_recoil}
\end{equation}
The remaining transverse-balance variables are defined as
\begin{align}
    J_{\mathrm{bal}}&=\left|\mathbf{p}_T^{\ell\ell}+\mathbf{p}_T^{\mathrm{miss}}\right|-p_T^{\ell\ell},\\
    m_T^{(p)}&=\sqrt{2p_T^{\ell\ell}p_T^{\mathrm{miss}}(1-\cos\Delta\phi_{\ell\ell,\mathrm{miss}})},\\
    m_T^{(E)}&=\sqrt{2E_T^{\ell\ell}p_T^{\mathrm{miss}}(1-\cos\Delta\phi_{\ell\ell,\mathrm{miss}})},
    \label{eq:idm_balance}
\end{align}
where $E_T^{\ell\ell}=\sqrt{m_{\ell\ell}^2+(p_T^{\ell\ell})^2}$.

To account for the imbalance between signal and background, their total training weights are rescaled to be equal. The signal benchmark contributions are also balanced, so that each benchmark receives the same total training weight. Within a benchmark, the relative VBF and s-channel normalization is preserved. This procedure allows the model to learn the kinematic structure of the smaller-cross-section benchmarks while retaining the physical production mixture at each hypothesis.

\begin{figure}[htp!]
    \centering
    \includegraphics[width=0.5\linewidth]{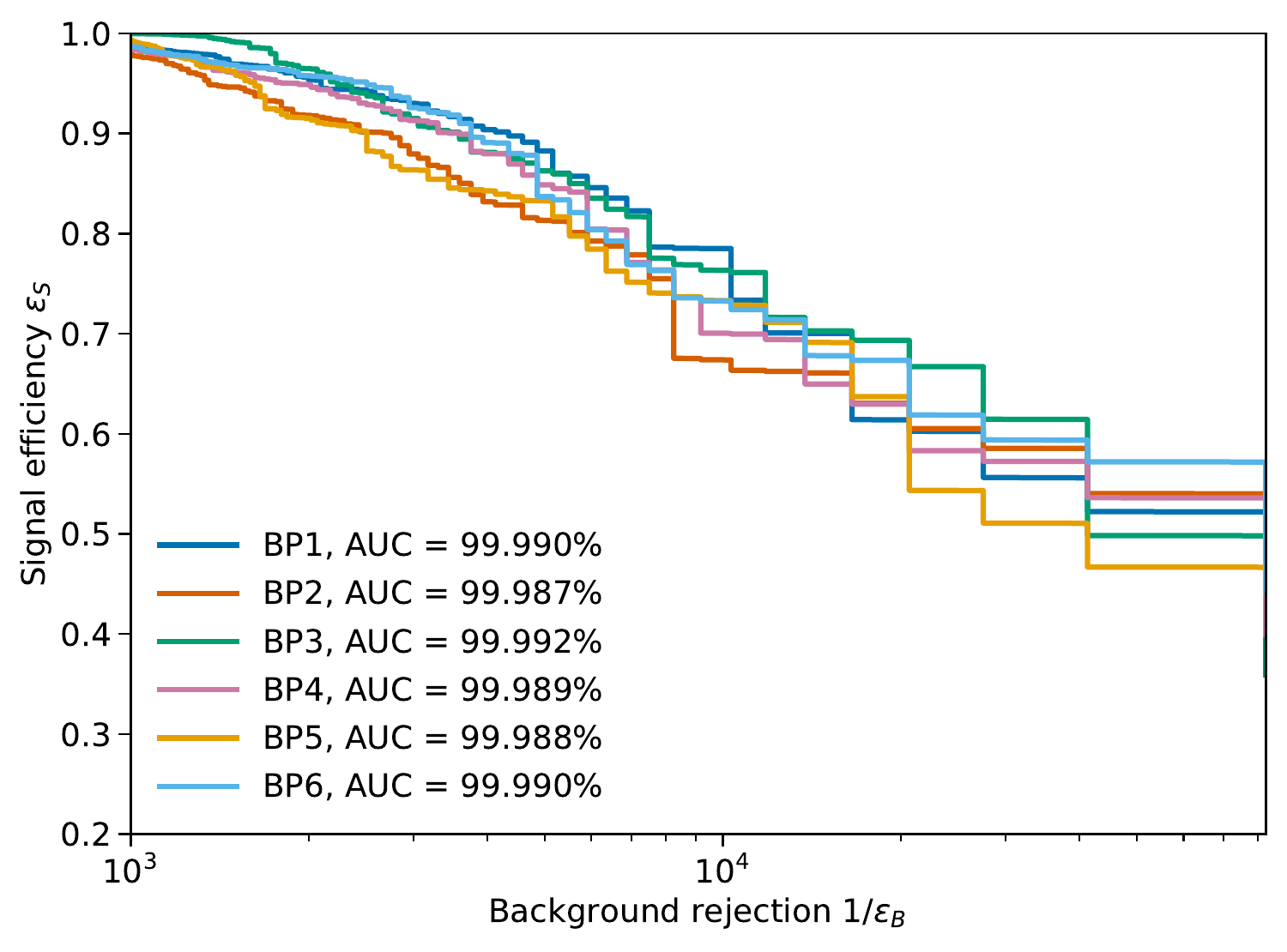}
    \caption{ROC curve for all benchmarks.}
    \label{fig:roc_all_benchmarks}
\end{figure}

\begin{figure}[htb!]
    \centering
     \includegraphics[width=0.49\textwidth]{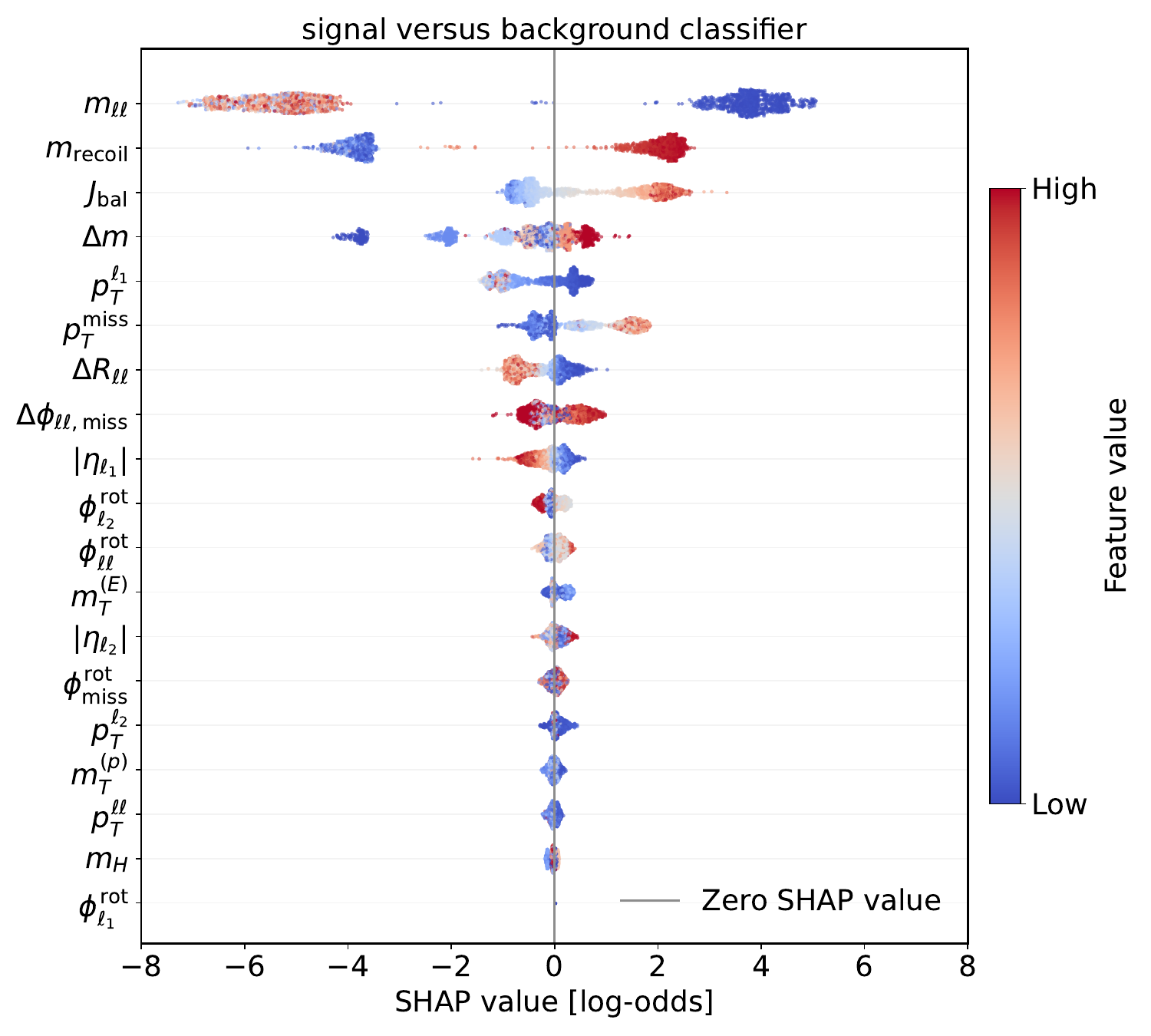}
   \includegraphics[width=0.49\textwidth]{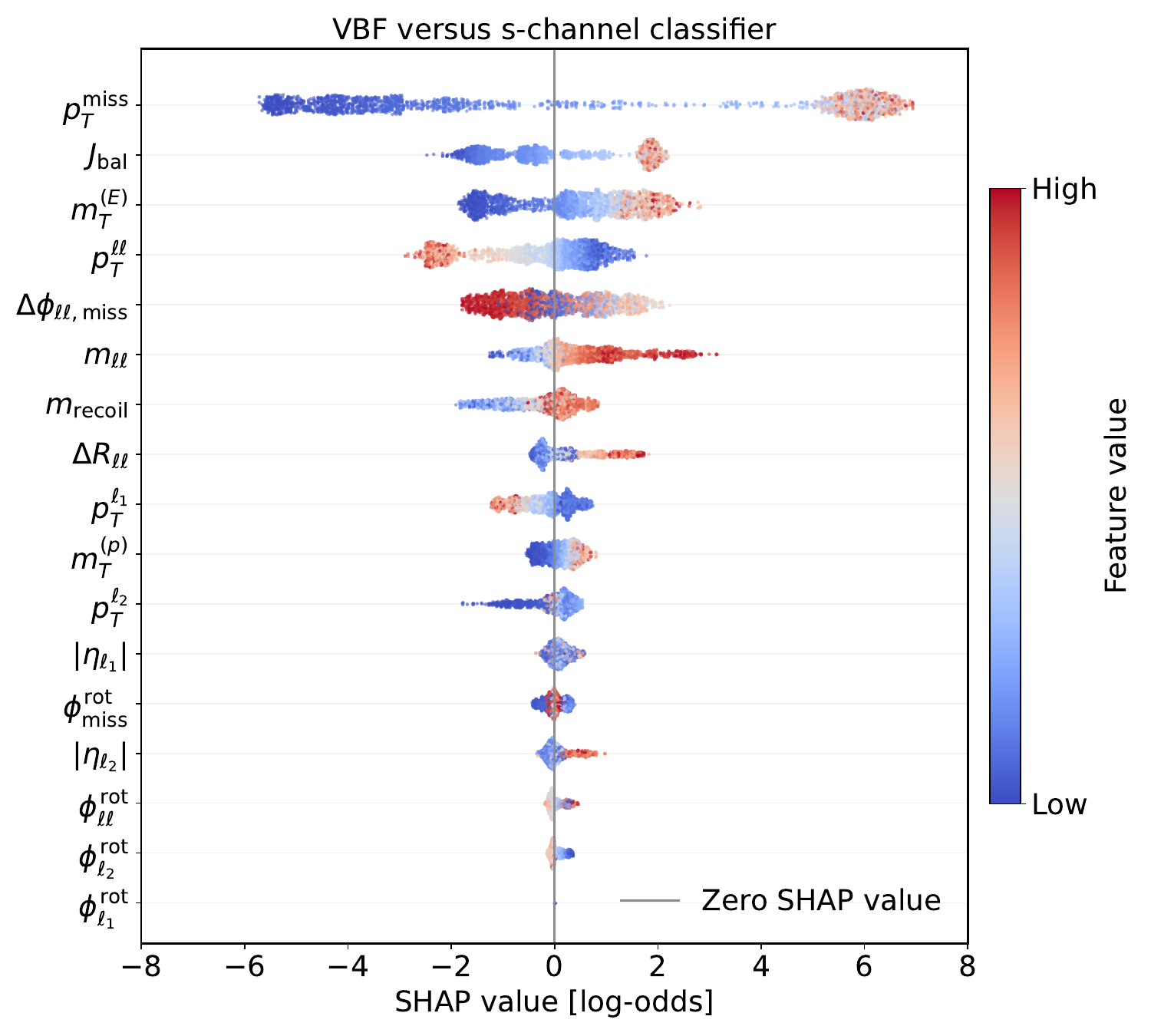}
    \caption{SHAP summary plots of the input features for signal-versus-background (left) and VBS-versus-s-channel (right) XGBoost classifiers. }
    \label{fig:idm_signal_importance}
\end{figure}

The hyperparameters used for both classifiers are listed in Table~\ref{tab:idm_bdt_parameters}. For the signal-versus-background classifier, training is allowed up to 1200 boosting iterations, with early stopping after 60 iterations without improvement in the validation logarithmic loss. Its weighted test area under the receiver operating characteristic curve (AUC) ranges from 99.98\% to 99.99\% across all training benchmarks. Figure \ref{fig:roc_all_benchmarks} shows the ROC curve for each benchmark. Figure \ref{fig:idm_signal_importance} (left) shows the relative features importance of the input features, as quantified by SHAP (SHapley Additive exPlanations) values.  The dilepton invariant mass is the highest-ranked feature according to the XGBoost gain importance, followed by the leading-lepton transverse momentum and the recoil mass. Including $m_{\ell\ell}$ is consistent with the final analysis strategy, since the likelihood is constructed from the classifier score rather than from a separate invariant-mass fit.

\begin{table}[htb!]
    \centering
    \begin{tabular}{lcc}
    \hline\hline
    Parameter & Signal/background & VBF/s-channel \\
    \hline
    \texttt{max\_depth} & 4 & 3 \\
    Learning rate & 0.03 & 0.03 \\
    \texttt{min\_child\_weight} & 5 & 5 \\
    \texttt{subsample} & 0.8 & 0.8 \\
    \texttt{colsample\_bytree} & 0.8 & 0.8 \\
    $L_2$ regularization & 2 & 2 \\
    $L_1$ regularization & 0 & 0 \\
    Maximum boosting rounds & 1200 & 1000 \\
    Early-stopping patience & 60 & 50 \\
    \hline\hline
    \end{tabular}
    \caption{XGBoost hyperparameters used in both classification tasks.}
    \label{tab:idm_bdt_parameters}
\end{table}

\begin{figure}
    \centering
    \includegraphics[width=0.45\textwidth]{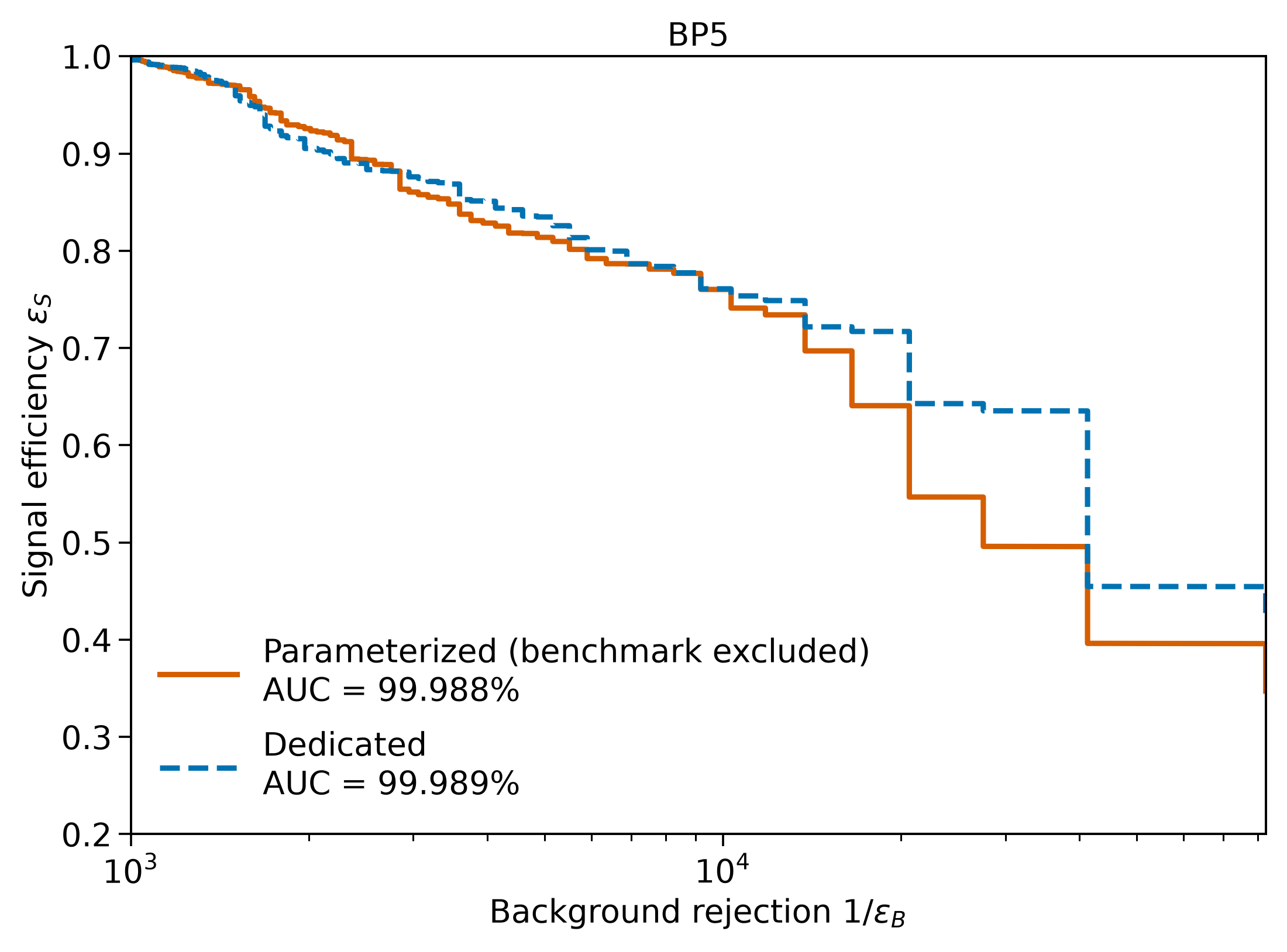}
    \includegraphics[width=0.45\textwidth]{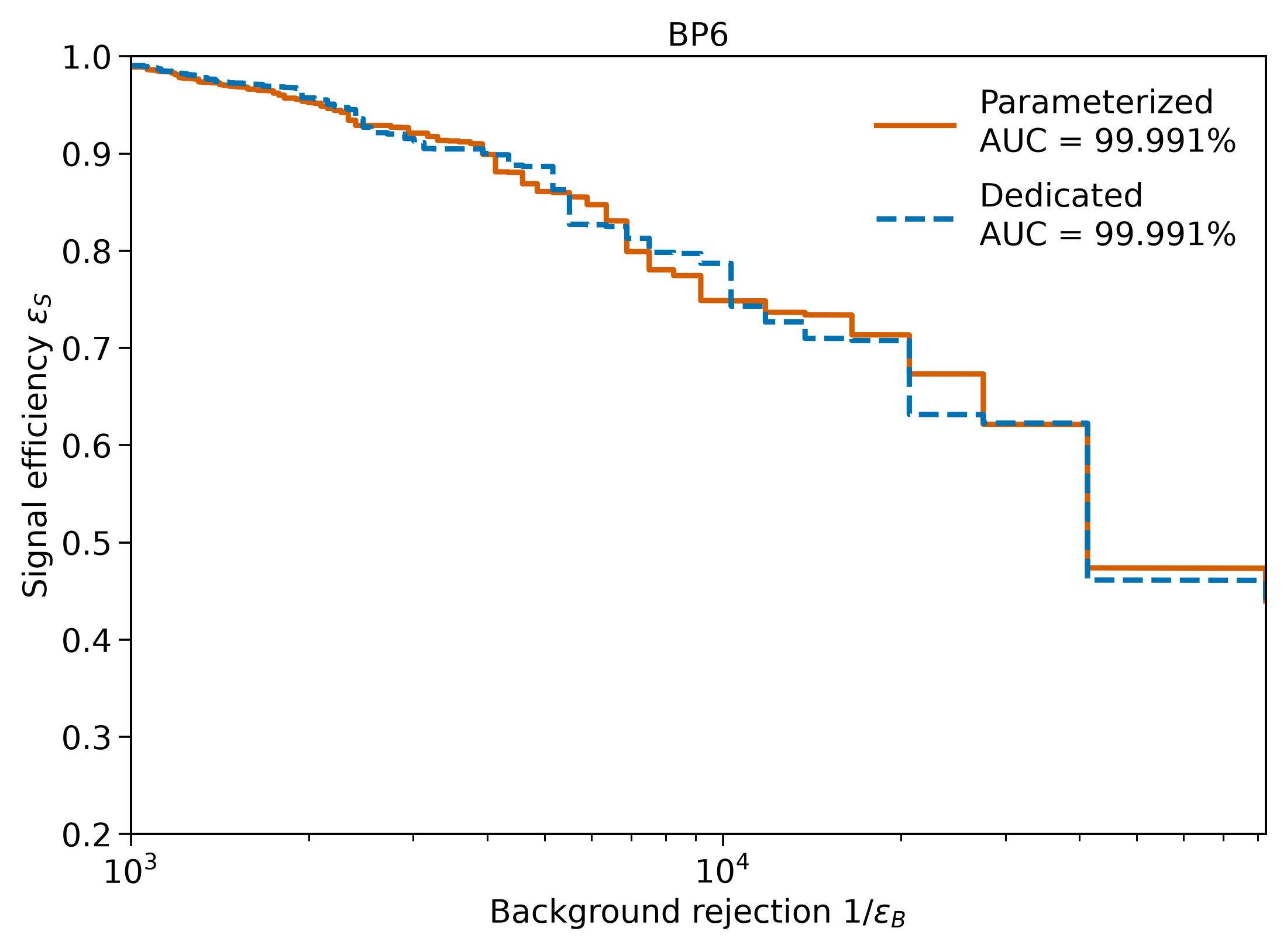}
    \caption{Comparison of ROC curve of benchmarks BP5 (left) and BP6 (right) on dedicated training (blue) and parameterized (orange) training. The benchmarks are excluded from the training. }
    \label{fig:parameterized}
\end{figure}

\begin{figure}
    \centering
    \includegraphics[width=0.5\linewidth]{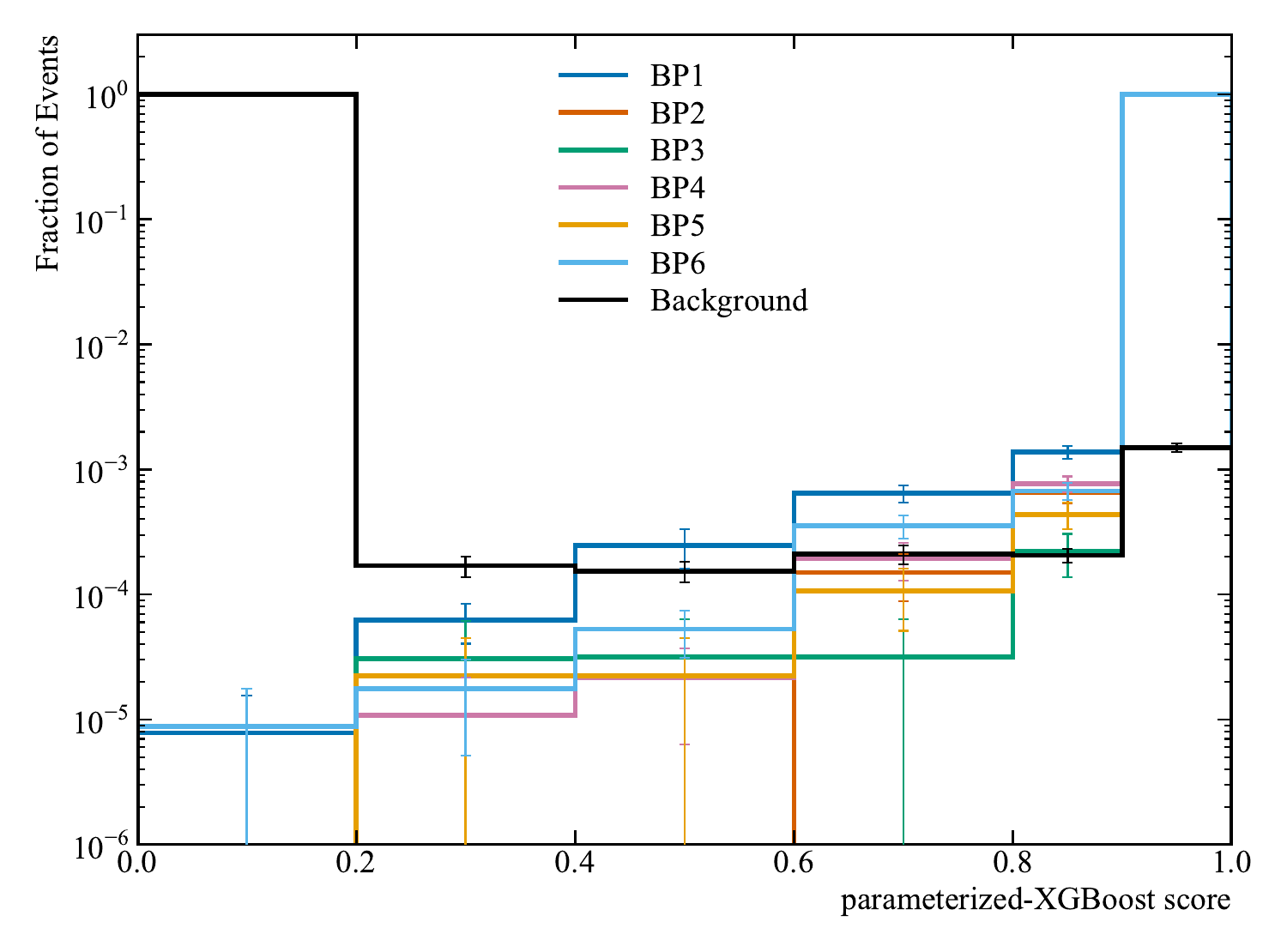}
    \caption{parameterized signal to background XGBoost score. For the BP5 and BP6 the parametrized XGBoost is used.}
    \label{fig:idm_signal_background_score}
\end{figure}

The ability of the parameterized classifier to generalize to untrained signal hypotheses is assessed in a separate benchmarks not included in training. The benchmarks at $\Delta m=33.18$ and 71.67 GeV are excluded from model fitting and early-stopping validation. The classifier is trained using only the hypotheses at $\Delta m=23.57$, 46.82, 61.51, and 87.88 GeV, with background conditioning restricted to the same training points. Its performance at the two excluded benchmarks is compared with that of dedicated classifiers trained directly at these hypotheses. For this comparison, the ROC curve is shown in Figure \ref{fig:parameterized}. The corresponding reductions are approximately 1.8\% and 1.0\%, respectively as summarized in Table \ref{tab:idm_interpolation}. Figure \ref{fig:idm_signal_background_score} shows the parameterized classifier score for background and all the benchmarks signal.

\begin{table}[htb!]
    \centering
    \begin{tabular}{lccc}
    \hline\hline
    $\Delta m$ [GeV] & Classifier & Weighted test AUC & $Z_{\mathrm{proxy}}$ \\
    \hline
    33.18 & Parameterized, benchmark excluded & 0.99981 & 6.90 \\
    33.18 & Dedicated & 0.99989 & 7.03 \\
    71.67 & Parameterized, benchmark excluded & 0.99956 & 13.87 \\
    71.67 & Dedicated & 0.99991 & 14.01 \\
    \hline\hline
    \end{tabular}
    \caption{Comparison of parameterized and dedicated classifiers at the two signal hypotheses excluded from parameterized training.}
    \label{tab:idm_interpolation}
\end{table}

\subsection{VBF-versus-s-channel XGBoost classifier}
\label{sec:idm_process_bdt}
The second classifier is designed to distinguish VBF from s-channel signal production. Although both processes lead to the selected dilepton-plus-missing-momentum signature, they exhibit different correlations among the visible and invisible systems. Separating these topologies allows their different signal-to-background compositions to be retained in the statistical analysis, allowing access to different couplings.

The classifier is trained using the BP1 signal samples only, with VBF events assigned label 1 and s-channel events label 0. Background events are not used in this training. The 17 reconstructed kinematic inputs listed in Table~\ref{tab:idm_bdt_inputs} are used, without the conditioning parameters $m_H$ and $\Delta m$. Physical event weights are rescaled so that the two production classes have equal total training weight. The production-classifier partition is generated independently of the signal-classifier partition.

\begin{figure}[htb!]
    \centering
    \includegraphics[width=0.75\linewidth]{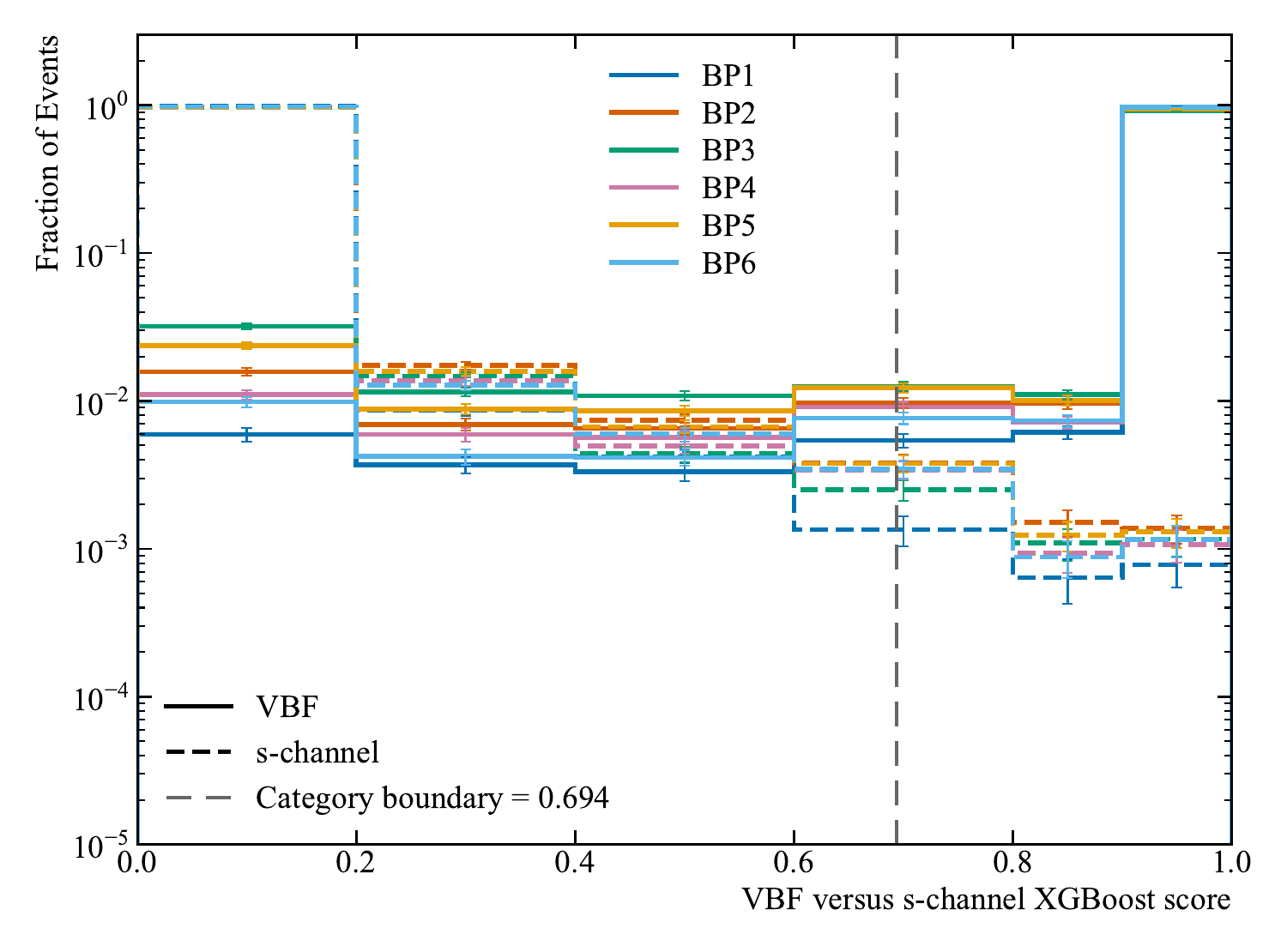}
    \caption{The production classier output score. The vertical black line corresponds to the category boundary used to assign events to each category.}
    \label{fig:idm_vbf_schannel_score}
\end{figure}

The model uses a maximum tree depth of three and the remaining hyperparameters listed in Table~\ref{tab:idm_bdt_parameters}. The output score shown in Figure \ref{fig:idm_vbf_schannel_score}, denoted by $D_{\mathrm{prod}}$, is used to define two mutually exclusive analysis categories:
\begin{equation}
    \begin{aligned}
        \mathcal{C}_{\mathrm{VBF}} &: D_{\mathrm{prod}}\geq0.694,\\
        \mathcal{C}_{s} &: D_{\mathrm{prod}}<0.694.
    \end{aligned}
    \label{eq:idm_categories}
\end{equation}

The threshold is selected on the BP1 validation subset by maximizing the difference between the VBF efficiency and the s-channel mis-classification rate. On the independent BP1 process-classifier test subset, the VBF efficiency is 98.30\%, while the s-channel mis-classification rate into the VBF category is 0.326\%.

The trained BP1 classifier is subsequently applied to the other signal benchmarks and to the background samples to determine their category membership. In the samples used to construct the limit templates, the VBF-category efficiency ranges from approximately 94.1\% to 98.5\% across the benchmarks, while the s-channel-category efficiency is approximately 99.6\% or higher. These efficiencies refer to the limit-template subsets and therefore differ slightly from the BP1 process-classifier test results quoted above.

The two outputs provide complementary information for the final analysis. The production score defines the event category, while the signal-versus-background score supplies the distribution fitted within that category. Background events assigned to the VBF category are treated as background in the fit.

\section{Statistical analysis}\label{sec:stat}

Following the event selection and machine learning categorization, a statistical interpretation is performed using the output score of the signal-versus-background classifier. Expected 95\% confidence level (CL) upper limits are determined separately for the VBF-enriched and s-channel-enriched orthogonal categories defined by the boundary $D_{\mathrm{prod}}=0.694$, as well as from their mutual combination. A background normalization uncertainty of 10\% is included in all three configurations.

Within each category, the parameterized signal-versus-background score, $D_{\mathrm{sig}}$, is used as the final observable. This procedure retains the different signal and background compositions of the two production topologies while exploiting the signal-discrimination information throughout the score distribution.

For each benchmark hypothesis, the background events are evaluated at the corresponding values of the conditioning parameters. Separate signal and background templates are then constructed in each production category from the simulated events. The templates use events from the signal-classifier test subset, with each sample rescaled to its expected yield at the full integrated luminosity. The computed limits use the parameterized classifier trained on all six benchmark hypotheses. 

The likelihood binning is optimized by scanning the classifier output score to identify the signal-enriched region that maximizes the expected Asimov significance. The highest-score bin is required to contain at least one expected background event, evaluated using the physical event weights. This requirement balances the signal sensitivity against the background population in the most discriminating region. The remaining score range is retained in the fit through additional bins.

As mentioned before, the statistical analysis is based on a binned likelihood implemented with the \texttt{pyhf} framework~\cite{Heinrich:2021gyp}. The parameter of interest is a nonnegative signal-strength modifier $\mu$, which scales the nominal VBF and s-channel signal cross-section simultaneously. For category $c$ and score bin $i$, the expected event yield is expressed as
\begin{equation}
    \nu_{ci}(\mu,\theta)=
    \mu\left(s_{ci}^{\mathrm{VBF}}+s_{ci}^{s}\right)
    +b_{ci}\,\kappa(\theta),
    \label{eq:idm_expected_bin_yield}
\end{equation}
where $s_{ci}^{\mathrm{VBF}}$ and $s_{ci}^{s}$ are the nominal signal predictions, and $b_{ci}$ is the total background prediction. The nuisance parameter $\theta$ represents the background normalization uncertainty fixed at 10\%. The combined likelihood is given by
\begin{equation}
    \mathcal{L}(\mu,\theta)=
    \prod_{c\in\{\mathrm{VBF},s\}}\prod_i
    \frac{\nu_{ci}(\mu,\theta)^{n_{ci}}
    e^{-\nu_{ci}(\mu,\theta)}}{n_{ci}!}
    \,\pi(\theta),
    \label{eq:idm_binned_likelihood}
\end{equation}
where $n_{ci}$ denotes the event count in each bin and $\pi(\theta)$ is the nuisance-parameter constraint. The category-only likelihoods are obtained by retaining the bins of the corresponding category.

The nuisance parameter is constrained by a standard normal distribution, and the normalization response is interpolated using the \texttt{normsys} modifier in \texttt{pyhf}. No independent background shape uncertainty is included.

The expected limits are evaluated using a background-only Asimov dataset, for which the event count in each bin is set to its nominal background expectation, $n_{ci}^{A}=b_{ci}$. The upper limit is determined using the asymptotic $\mathrm{CL}_{s}$ prescription~\cite{Read:2002hq} with a one-sided profile-likelihood statistic and the physical constraint $\mu\geq0$. The median expected limit, $\mu_{95}$, is obtained at $\mathrm{CL}_{s}=0.05$, together with the expected $\pm1\sigma$ and $\pm2\sigma$ intervals. 

For each benchmark, the theoretical signal cross-section is defined as the sum of the VBF and s-channel contributions,
\begin{equation}
    \sigma_{\mathrm{thoe}}=
    \sigma_{\mathrm{VBF}}^{\mathrm{theo}}+
    \sigma_s^{\mathrm{theo}}.
    \label{eq:idm_nominal_signal_cross_section}
\end{equation}
The relative normalization of these contributions is fixed to its benchmark value. The cross-section upper limit is therefore obtained from the signal-strength limit according to
\begin{equation}
    \sigma_{95}=\mu_{95}\,\sigma_{\mathrm{thoe}}.
    \label{eq:idm_cross_section_limit}
\end{equation}
Any decay branching fractions already included in the generator cross-sections remain included in the limited quantity; no additional branching-fraction correction is applied in this fit.

\begin{figure}[htb!]
    \centering
    \includegraphics[width=0.7\textwidth]{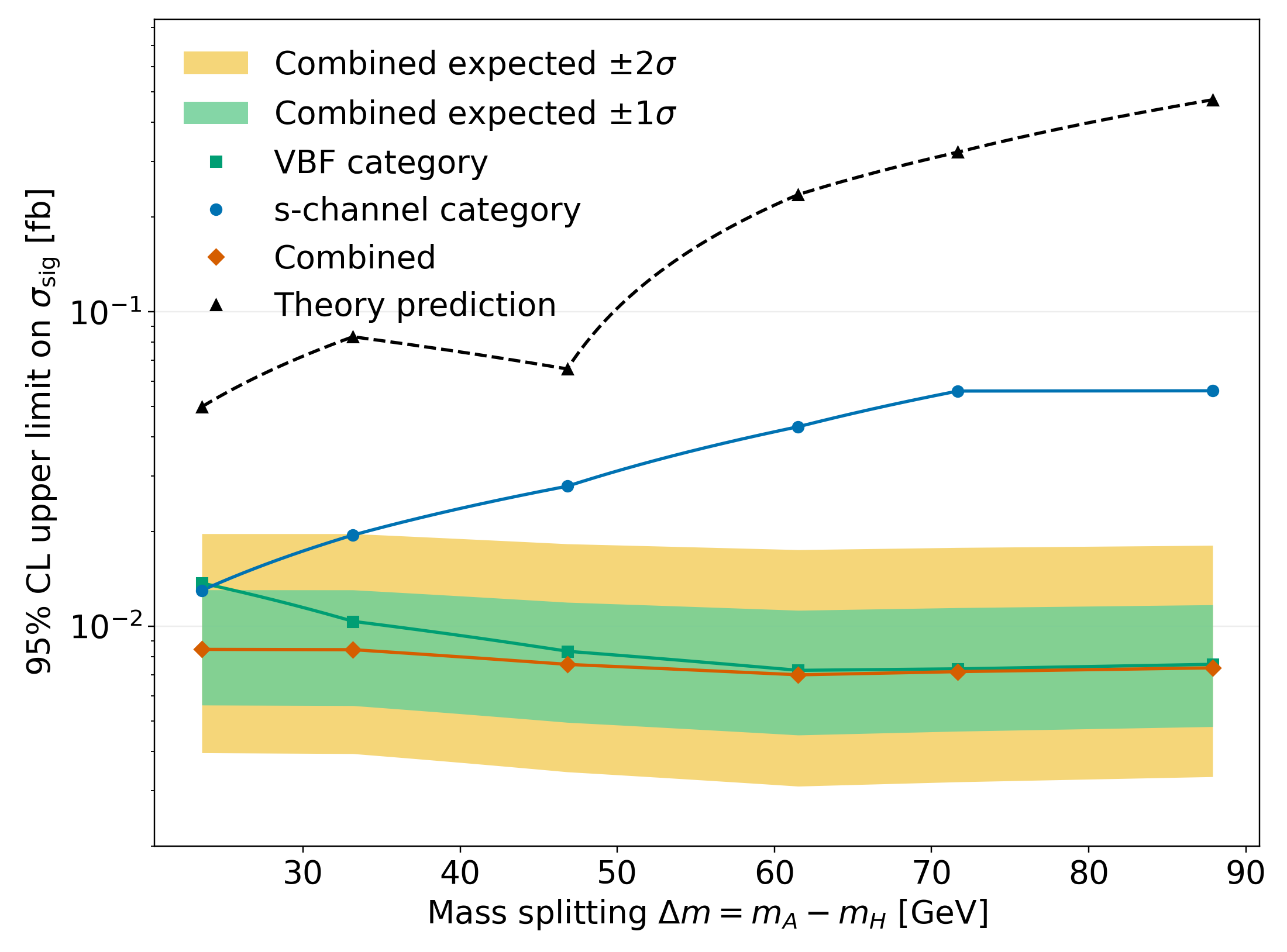}
    \caption{Expected 95\% CL upper limits on the total signal cross-section as a function of the mass splitting, obtained using the VBF-enriched category, the s-channel-enriched category, and their combination. The bands indicate the expected $\pm1\sigma$ and $\pm2\sigma$ intervals. The theoretical signal prediction is shown for comparison.}
    \label{fig:idm_expected_limits}
\end{figure}

Figure~\ref{fig:idm_expected_limits} shows the expected cross-section limits and their corresponding $\pm1\sigma$ and $\pm2\sigma$ intervals as a function of the mass splitting. The combined median limit changes across the sampled hypotheses, decreasing from approximately $7.82\times10^{-3}$ fb at the smallest splitting to about $6.1\times10^{-3}$ fb at the largest splittings. These points correspond to distinct benchmarks with different masses and production mixtures.

The VBF-enriched category provides the dominant sensitivity for the benchmark considered here. This behavior follows from the larger VBF signal contribution for most benchmarks and the smaller background yield in this category. For BP1, the nominal VBF and s-channel cross-sections are 0.4146 and 0.05707 fb, respectively. The VBF-category limit is $0.00615$ fb, whereas the s-channel-category limit is $0.07142$ fb. Their combination gives $0.00609$ fb, representing an improvement of approximately 0.9\% relative to the VBF category alone. The small improvement demonstrates that the BP1 sensitivity is driven primarily by the VBF-enriched sample.

The s-channel contribution becomes more important when the two production rates are similar. For BP3, the nominal cross sections are $0.02534~\mathrm{fb}$ for the s-channel and $0.02451~\mathrm{fb}$ for VBF. Combining the categories improves the median expected limit from $0.01034$ fb for the VBF category alone to $0.00782$ fb, corresponding to an improvement of approximately 24\%. The category combination therefore retains useful information about both production topologies, even though the VBF-enriched category dominates for the larger VBF fractions.

\section{Conclusions}\label{sec:conclusion}



We find that, for a $3$~TeV MuC, the cut-based selection already yields a counting sensitivity of $Z \simeq 4.3$ for direct production and $Z \simeq 13.7$ for the VBF-enhanced channel at an integrated luminosity of $1~\text{ab}^{-1}$ for our first benchmark. The multivariate analysis leads to expected upper limits of $(6$--$8)\times 10^{-3}$~fb on the total signal cross section across the benchmarks, well below the predicted rates. The VBF-enriched category drives the sensitivity in most of the parameter space, while the $s$-channel category improves the combined limit by up to about $24\%$ when the two production rates are comparable. These results show that a high-energy MuC can test the compressed inert spectrum in a region that is difficult to access at the LHC, and that the combination of production-mode categorisation with parameterised multivariate classifiers is an efficient strategy for such soft-lepton signatures.

\section*{Acknowledgments}
MB, AH, and LR would like to thank their collaborators for the fruitful discussions and valuable suggestions throughout this work.

\bibliographystyle{JHEP}
\bibliography{bibliography}

\end{document}